\documentclass[sigconf]{acmart}
\usepackage{booktabs} 
\usepackage{float}
\usepackage{multirow}
\usepackage{graphicx}
\usepackage{array}
\usepackage{xcolor}
\usepackage{soul}
\usepackage{wrapfig}
\usepackage{subcaption}

\newcolumntype{P}[1]{>{\centering\arraybackslash}p{#1}}
\newcolumntype{M}[1]{>{\centering\arraybackslash}m{#1}}
\AtBeginDocument{%
  \providecommand\BibTeX{{%
    \normalfont B\kern-0.5em{\scshape i\kern-0.25em b}\kern-0.8em\TeX}}}
    
\copyrightyear{2022} \acmYear{2022} \setcopyright{acmlicensed}\acmConference[CHI '22]{CHI Conference on Human Factors in Computing Systems}{April 29-May 5, 2022}{New Orleans, LA, USA} \acmBooktitle{CHI Conference on Human Factors in Computing Systems (CHI '22), April 29-May 5, 2022, New Orleans, LA, USA} \acmPrice{15.00} \acmDOI{10.1145/3491102.3517563} \acmISBN{978-1-4503-9157-3/22/04}

\begin{document}
\setlength{\tabcolsep}{1pt}
\title[An Understanding of Multimodal and Parasocial Interactions in ASMR videos]{Close-up and Whispering: An Understanding of Multimodal and Parasocial Interactions in YouTube ASMR videos}

\author{Shuo Niu}
\email{shniu@clarku.edu}
\orcid{https://orcid.org/0000-0002-8316-4785}
\affiliation{%
  \institution{Clark University}
  \streetaddress{950 Main St.}
  \city{Worcester}
  \state{MA}
  \country{USA}
  \postcode{01610}
}

\author{Hugh S. Manon}
\email{hmanon@clarku.edu}
\affiliation{%
  \institution{Clark University}
  \streetaddress{950 Main St.}
  \city{Worcester}
  \state{MA}
  \country{USA}
  \postcode{01610}
}

\author{Ava Bartolome}
\email{abartolome@clarku.edu}
\affiliation{%
  \institution{Clark University}
  \streetaddress{950 Main St.}
  \city{Worcester}
  \state{MA}
  \country{USA}
  \postcode{01610}
}

\author{Nguyen B. Ha}
\email{joha@clarku.edu}
\affiliation{%
  \institution{Clark University}
  \streetaddress{950 Main St.}
  \city{Worcester}
  \state{MA}
  \country{USA}
  \postcode{01610}
}

\author{Keegan Veazey}
\email{kveazey@clarku.edu}
\affiliation{%
  \institution{Clark University}
  \streetaddress{950 Main St.}
  \city{Worcester}
  \state{MA}
  \country{USA}
  \postcode{01610}
}
\renewcommand{\shortauthors}{Shuo Niu, et al.}

\begin{abstract}
ASMR (Autonomous Sensory Meridian Response) has grown to immense popularity on YouTube and drawn HCI designers' attention to its effects and applications in design. YouTube ASMR creators incorporate visual elements, sounds, motifs of touching and tasting, and other scenarios in multisensory video interactions to deliver enjoyable and relaxing experiences to their viewers. ASMRtists engage viewers by social, physical, and task attractions. Research has identified the benefits of ASMR in mental wellbeing. However, ASMR remains an understudied phenomenon in the HCI community, constraining designers' ability to incorporate ASMR in video-based designs. This work annotates and analyzes the interaction modalities and parasocial attractions of 2663 videos to identify unique experiences. YouTube comment sections are also analyzed to compare viewers' responses to different ASMR interactions. We find that ASMR videos are experiences of multimodal social connection, relaxing physical intimacy, and sensory-rich activity observation. Design implications are discussed to foster future ASMR-augmented video interactions.
\end{abstract}



\maketitle

\section{Introduction}
Autonomous Sensory Meridian Response (ASMR) is a phenomenon usually experienced as tingling sensations in the crown of the head in response to a range of audio-visual triggers such as whispering, tapping, and hand movements \cite{PoerioMoreThanAFeeling}. ASMR videos incorporate audio, touch, taste, observation, and roleplay effects to deliver enjoyable and relaxing feelings. Over the past decade, the creation culture on YouTube has attracted numerous ASMR creators (known colloquially by users as ``ASMRtists'') to design a wide array of tingle-inducing sounds and actions to intentionally induce ASMR feelings \cite{ManonASMR, AndersenShiveriesASMR}. ASMRtists have also leveraged ASMR videos to connect to the viewers and build online ASMR communities \cite{SmithAffectIntimacy, AndersenShiveriesASMR}. A typical YouTube ASMR video may feature an ASMRtist whispering to the viewer, roleplaying personal attention such as massages or haircuts, making crisp sounds, or engaging in various slow and repetitive movements \cite{BarrattASMRFlowLike}. YouTube hosted more than 5.2 million ASMR videos in 2016 and 13 million in 2019, and the searches for ASMR grew over 200\% in 2015 and are consistently increasing \cite{MooneyASMRYouTubeTrend, StearsThreeThingASMR}. Remarkably ``ASMR'' is among the top five  YouTube search queries globally and in the US, with a search volume of more than 14 million\footnote{https://ahrefs.com/blog/top-youtube-searches/}. 
\par
In Human-Computer Interaction (HCI), experience-centered design requires researchers to capture and analyze the experiences generated from interaction and adopt the understanding of these experiences in design practices \cite{HassenzahlExperienceDesign}. ASMR is a unique experience insofar as only some users experience the ``tingles'' as a response to particular triggers, and the same trigger may have different effects on different people \cite{ManonASMR,FredborgASMRPersonality,  McErleanIndividualVariationASMR, PoerioMoreThanAFeeling}. Over the years, ASMRtists developed highly stylized and conventionally patterned ASMR videos to engage their viewers, and as a way to enhance affect and intimacy \cite{AndersenShiveriesASMR, ZappavignaASMRRolePlay}. Prior research on ASMR has focused on characterizing ASMR triggers \cite{BarrattASMRFlowLike, FredborgMindfulness} or understanding ASMR interactions through qualitative video analysis \cite{ZappavignaASMRRolePlay, AndersenShiveriesASMR, SmithAffectIntimacy}, user surveys \cite{LiuASMRTriggering, PoerioMoreThanAFeeling}, and brain imaging \cite{SmithASMRNetwork}. Most studies described YouTube ASMR videos primarily as roleplays \cite{AhujaFeelsGoodTobeMeasured, ZappavignaASMRRolePlay, StarrASMRSexuality} or as a single video type with a mixture of ASMR triggers \cite{SmithAffectIntimacy, PoerioMoreThanAFeeling}. However, little data-driven research has been conducted to categorize the wide variety of ASMR experiences developed by YouTube ASMRtists. YouTubers create ASMR videos with or without elements of social interaction, using roleplays or simply manipulating objects, and position themselves up-close or distant from the viewer. A macro understanding of the delivery mechanism and experience patterns in YouTube ASMR videos will help technology and service designers explore ways to integrate ASMR and assess its effects on the user experience. Since experience seekers need different triggers to acquire ASMR sensations, a quantitative overview of common ASMR interaction modalities will indicate what ASMR interactions may work for more users.
\par
In this study, we collect a large number of ASMR videos and perform a mixed-method analysis to obtain an overview of ASMR interactions and experiences. This work analyzes 2663 ASMR videos collected from YouTube to examine the multimodal interactions and the ways ASMR performers para-socially attract the viewers. We focus on intentional ASMR videos -- videos with ``ASMR'' labels in which a variety of triggers are purposefully displayed by the performer -- to understand ASMRtists' common approaches to trigger ASMR experiences. Prior work identified visual, audio, touch, taste, and scenario-based ASMR triggers \cite{RichardBrainTingles, ZappavignaASMRRolePlay, SmithFuctionalTriggers, FredborgASMRPersonality}. By interacting with ASMR videos, viewers are able to experience a simulation of intimacy with the video performer through ``parasocial interactions'' \cite{WohnParasocialInteraction} -- a one-sided intimacy experienced by a viewer through repeated encounters with a figure on screen. In parasocial relationships, video performers develop and manage three types of attractiveness -- \textit{social attraction}, \textit{physical attraction}, and \textit{task attraction} \cite{RubinDevelopmentofParasocialInteraction}. We quantify the manners in which the ASMRtists socialize with the viewer (social attraction), the camera proximity of the ASMRtists in the videos (as an alternative for physical attraction), and purposeful activities performed by the ASMRtists (task attraction). This work addresses three main research questions:
\begin{itemize}
  \item RQ1: How are various interaction modalities employed in YouTube ASMR videos?
  \item RQ2: How do YouTube ASMRtists design parasocial attractiveness through multimodal interactions?
  \item RQ3: How do different multimodal interactions and parasocial attractions affect the expression of viewers' feelings in the comments?
\end{itemize}

\begin{figure}[ht]
    \centering
    \scalebox{1}{
       \includegraphics[width=\linewidth]{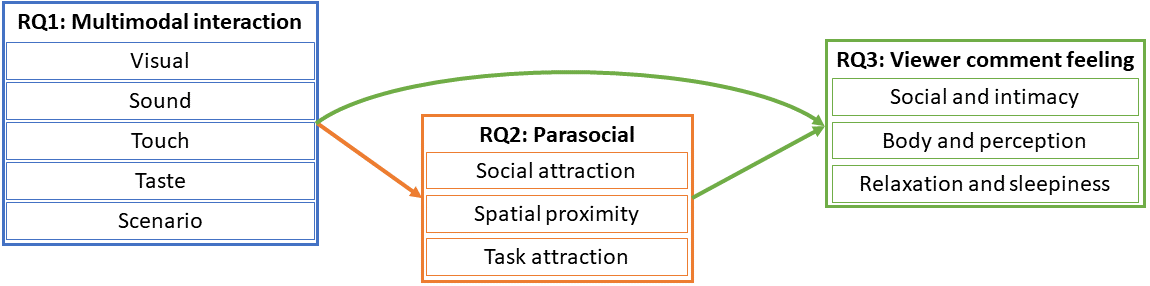}
    }
    \caption{The structure of the research questions}
    \Description{RQ1 examines visual sound, touch, taste, and scenario modalities. RQ2 examines how interaction modalities predict social attraction, spatial proximity, and task attraction. RQ3 examines how multimodal interactions and parasocial attractions predict social and intimacy, body and perception, and relaxation and sleepiness in viewer comments.}
    \label{fig:researchquestion}
\end{figure}

Figure \ref{fig:researchquestion} illustrates the structure of the research questions. RQ1 provides an overview of multimodal interactions in YouTube ASMR videos to inform designs with common ASMR performing methods. RQ2 focuses on understanding the patterns of parasocial attractiveness through multimodal interactions. We summarize the experiences delivered by YouTube ASMR videos and identify the associated interaction modalities. RQ3 utilizes viewers' comments to infer how different multimodal interactions and parasocial attractions affect viewers' social, perceptual, and relaxation feelings. We first use grounded-theory approaches to identify subcategories of interaction modalities and parasocial attractions. Then the codebook is translated into a questionnaire task. The annotation tasks were completed by participants recruited from Amazon Mechanical Turk (MTurk). We perform statistical analysis to address the research questions. 
\par
The development of multimodal interactions depends on the natural integration patterns that typify the combined use of different input modes \cite{OviattMythMultimodalInteraction}. Understanding diverse interaction modalities through analyzing extensive video data inform different ways to incorporate ASMR in technology design. Our results indicate social attractions are enhanced by combining multiple ASMR interaction modalities. Most ASMRtists use the closeup camera proximity as a means of building physical attractiveness. ASMRtists emulate physical closeness through microphonically-amplified whispering, manipulating objects, virtually ``touching'' the viewer, and making mouth noises and microphone-jostling sounds near the camera or the microphone. Many ASMR videos do not involve purposeful tasks and are not roleplays. Tasks used in non-roleplay videos include soft and routine activities such as performing medical or cosmetic treatments, eating and drinking, and demonstrating mundane daily activities. The ASMR experiences delivered by YouTube ASMRtists can be described as three experience patterns: multimodal social connection, relaxing physical intimacy, and observation of sensory-rich activities. This work aims to inspire future technologies and services to incorporate ASMR triggers to design ASMR-augmented relaxing or intimate experiences. 

\section{Background}
\subsection{ASMR Videos on YouTube}
The now widely-adopted term ``Autonomous Sensory Meridian Response'' (ASMR) was coined in 2010 to describe a sensory phenomenon that usually involves the sensation of tingling as a response to certain audio-visual stimuli \cite{BarrattSensoryDeterminants}. Common ASMR videos show intentional or unintentional gentle interactions such as speaking softly, playing or brushing hair, moving hands, and tapping or scratching surfaces \cite{PoerioMoreThanAFeeling, FredborgASMRPersonality}, which may trigger a low-grade euphoria response and tingling sensations on the viewer's  head and spine \cite{AhujaFeelsGoodTobeMeasured}. The ASMR trend on social media began with a Yahoo group sharing personal experiences of head tingles when watching specific kinds of videos \cite{CheadleASMR, AndersenShiveriesASMR}. Those original videos were dubbed ``unintentional'' ASMR, and involved real-world scenarios such as doctor's office examinations and suit fittings, captured for some non-ASMR purpose and uploaded, but subsequently re-contextualized for their ASMR tingle-triggering properties \cite{ManonASMR, GallagherAetheticsOfASMR}. Afterward, creators made numerous ``intentional'' ASMR videos on YouTube in which ASMRtists purposefully use visual and sound stimuli and scripted roleplays to induce ASMR experiences \cite{AhujaFeelsGoodTobeMeasured, MaddoxASMRCommunity}. In 2019, there were 13 million ASMR videos on YouTube \cite{StearsThreeThingASMR}. Popular ASMRtists such as GentleWhispering ASMR, SAS-ASMR, and Gibi ASMR, have millions of subscribers, and their videos attracted millions of views and generate considerable revenue for the creators \cite{ZappavignaASMRRolePlay, StearsThreeThingASMR}.
\par
Despite the popularity of this emerging video genre on YouTube, studies found ASMR triggers do not work for everyone, and some individuals only experience the tingles with very precise, idiosyncratic triggers \cite{BarrattASMRFlowLike, PoerioMoreThanAFeeling}. ASMR was found to be associated with specific personality traits of individual viewers and to vary from person to person \cite{FredborgASMRPersonality, McErleanIndividualVariationASMR}. Users' diverse needs triggering effects drive ASMR consumers to constantly search for videos with the keyword ``ASMR.'' \cite{AndersenShiveriesASMR} In turn, the YouTube culture of creativity and participation \cite{Burgess2018YouTube:Culture} encourages ASMRtists to make numerous ASMR videos to satisfy ASMR experience-seekers' diverse needs. The growing trend of ASMR creation and consumption drew researchers' attention. Prior studies focused on understanding the sensational effects through interviewing ASMR viewers (e.g., \cite{LiuASMRTriggering, BarrattASMRFlowLike, PoerioMoreThanAFeeling}) or scanning brain images (e.g., \cite{SmithASMRNetwork}). Some studies examined the creator-viewer interactions and the digital intimacy through qualitative analyses of a few viral videos (e.g., \cite{SmithAffectIntimacy, ZappavignaASMRRolePlay, MaddoxASMRCommunity}).
\par
Although the diversity of ASMR triggers is widely noted, there is little analysis of large video data sets to explain the creation practices employed by ASMRtists. ASMR can be induced with virtual face-to-face interactions \cite{BarrattASMRFlowLike, StarrASMRSexuality} or simply by manipulating objects without showing performers' faces \cite{Nissinen}. Some ASMR videos consist of constant soft speaking while others involve object manipulation without talking \cite{LapinskaNoTalking}. Some ASMR videos pretend to touch the viewers in roleplays while others perform massages on a second person who is also visible in the video \cite{ZappavignaASMRRolePlay}. ASMR sensations can emerge both in response to food consumption videos \cite{TangMukbang} and to videos showing a person studying quietly \cite{LeeStudyWithMe}. A quantitative analysis of extensive videos will help HCI designers discern the significance of different ASMR interactions and experiences. First, since an ASMR trigger may or may not induce ASMR experiences, an overview of common ASMR interaction modalities and experiential patterns will help technology designers to incorporate ASMR triggers that are effective for a broader range of users. Second, recent research noted ASMR is not just a sensory experience; it is also a kind of mediated intimacy offered by ASMRtists to deliver a sense of social connection \cite{AndersenShiveriesASMR, SmithAffectIntimacy, ZappavignaASMRRolePlay}. However, there is limited understanding of how such social experiences are commonly constructed and their relationships with trigger interactions and social settings. Last, understanding viewers' social, perceptual, and relaxation feelings will help technology designers understand the possible effects of different ASMR experiences.

\subsection{Multimodal Interactions in ASMR videos}
Researchers have examined various triggers in ASMR videos to understand this emerging media form and its physiological effects. Richard summarizes ASMR stimuli as audio, touch, visual, and scenario triggers \cite{RichardBrainTingles}. Smith et al. examined people's responses to five trigger types, including watching, touching, repetitive sounds, simulations, and mouth sounds \cite{SmithFuctionalTriggers}. Zappavigna explored ASMR roleplays and found linguistic, visual, and aural resources are used to create a sense of co-presence with the performer \cite{ZappavignaASMRRolePlay}. This study explores common interactions performed by YouTube ASMRtists. HCI researchers have identified vision, hearing, touch, smell, and taste as five sensing modalities to embody interactions \cite{SharmaMultimodalInterface, YuanMultiSensorial}. Grounded in multimodal interaction theories \cite{SharmaMultimodalInterface, YuanMultiSensorial} and trigger modalities identified in the literature \cite{RichardBrainTingles}, RQ1 explores \textit{visual}, \textit{sound}, \textit{touch}, \textit{taste}, and \textit{scenario} as five interaction modalities in YouTube ASMR videos. \textit{Visual interactions} describe how the performers present themselves and the trigger objects. We look into visual settings such as ASMRtists showing themselves in front of the camera, performing slow activities, or simply showing hands manipulating ASMR trigger objects. \textit{Sound interactions} refer to what types of hearing triggers are made by the ASMRtists. This modality seeks to capture sounds like human speaking, tapping or scratching objects, or various sound effects produced by interacting with the microphone. \textit{Touch interactions} examine how ASMRtists stimulate haptic feelings. For this modality, we observe how ASMRtists use their hands to interact with themselves, physical objects, the camera, or another person in the video. \textit{Taste interactions} investigate whether or not the ASMRtists eat food in the video. And finally, \textit{scenario triggers} describe the simulated situation and environment in roleplay videos, such as haircuts, eye exams, or other dramatized forms of assistance from the figure on screen \cite{RichardBrainTingles}. 

\subsection{Parasocial Attractions in ASMR Videos}
A large community of YouTube creators designate themselves as ``ASMRtists'' by regularly creating and uploading ASMR videos \cite{AndersenShiveriesASMR}. The pseudo-interactive nature of the videos engenders a sort of intimacy with the video creator \cite{HuVideoSharingCommunity, Cannell2018YouTubeInfluencer, NiuSHWM}. The essentially one-sided intimacy with the video performers, generated by a \textit{``conversational give-and-take,''} is defined as a parasocial relationship, and the interactions users have with videos that generate parasocial relationships are called \textit{``parasocial interactions.''} \cite{HortonParasocialInteraction, HartmannParasocialInteractionWellbeing} Parasocial relationships and interactions have been widely found in the interactive reponses of viewers to a TV or social media figure, which affect viewers socially and emotionally. Video-watching may lead some viewers to imagine themselves interacting with the performer \cite{GileParasocialInteraction}. Studies found parasocial relationships can provide social support and shield against the effects of exclusion and loneliness \cite{HartmannParasocialInteractionWellbeing, NiuSHWM}. YouTube users mostly focus their investments (of time, of energy) in parasocial interactions with the video creator, rather than building a friend and community network like other social media \cite{PreeceYouTubeCommunity, NiuTeamTrees}. ASMR videos can be seen as a unique form of parasocial interactions offered by ASMRtists \cite{Burgess2018YouTube:Culture}. Klausen described ASMR as a ``para-haptic interactional'' relation with the ASMRtists while obtaining a form of presence and intimacy \cite{KlausenSafeandSound}. Zappavigna also considered ASMR videos as a construction of the interactive context in which viewers feel co-presence with the performer \cite{ZappavignaASMRRolePlay}. Smith argued that in ASMR videos, affective experiences are intentionally construed and strategically heightened \cite{SmithAffectIntimacy}. However, due to a dearth of quantitative analysis of parasocial interactions in ASMR videos, it is unknown how prevalent the social experience is, or what general approaches are best used to deliver such one-sided intimacy.
\par
This work analyzes the associations between parasocial interactions and interaction modalities to explore the patterns of social and intimate experiences designed by ASMRtists. The parasocial interaction theory suggests that video performers develop \textit{social}, \textit{physical}, and \textit{task} attractions to engage viewers and establish parasocial relationships \cite{RubinDevelopmentofParasocialInteraction, KurtinDevelopmentofParasocilIntearctionYouTube}. \textit{Social attraction} refers to the degree to which one feels they would like to befriend the television  or media persona \cite{KurtinDevelopmentofParasocilIntearctionYouTube}. In ASMR, performers may simulate conversational scenarios and socialize with the viewers through their soft vocal and bodily interactions \cite{ZappavignaASMRRolePlay, KlausenSafeandSound}. Creators use ASMR videos to attract patrons and build network of fans \cite{MaddoxASMRCommunity}. \textit{Physical attraction} refers to how video performers appeal to the viewer physically \cite{KurtinDevelopmentofParasocilIntearctionYouTube}. In this study, we measure camera spatial proximity as a vector of physical attraction since it is difficult to quantify the attractiveness of performers' physical appearance. ASMRtists tend to perform body and hand movements very near the camera \cite{ZappavignaASMRRolePlay, StarrASMRSexuality}. ASMRtists may themselves appear in the video either in close up to show intimacy \cite{SmithAffectIntimacy, AndersenShiveriesASMR, HarperBodilyPleasure}, or they may exclusively show trigger objects on screen without showing their faces \cite{RichardBrainTingles}. \textit{Task attraction} describes how ably, credibly, or reliably a performer can complete a task \cite{KurtinDevelopmentofParasocilIntearctionYouTube}. ASMRtists perform different tasks such as professional treatments \cite{LapinskaFixingYou, ZappavignaASMRRolePlay, AhujaFeelsGoodTobeMeasured}, mundane activities like putting on makeups or sorting cards \cite{FredborgASMRPersonality, DevendorfASMRMundane, LeeStudyWithMe}, or less meaningful activities like cutting soap or tapping their fingernails on objects \cite{GallagherAetheticsOfASMR}. RQ2 uses the parasocial attraction framework identified in \cite{RubinDevelopmentofParasocialInteraction} to examine the patterns of social, proximity, or task-observing experiences.

\subsection{ASMR Experiences and Benefits}
ASMR experiences are touted by many as promoting calm and relaxed feelings \cite{PoerioASMRRelaxation, KovacevichASMRMusicRelaxing, MoralesASMRReappraisal} and are associated with positive affect and a sense of interpersonal connection \cite{PoerioMoreThanAFeeling, KlausenSafeandSound, SmithAffectIntimacy}. Barratt and Davis found that ASMR combines positive feelings, relaxation, and tingling sensation of the skin and provides temporary relief from depression \cite{BarrattASMRFlowLike}. Smith et al. analyzed neuroimages during ASMR tingles and found ASMR was associated with a blending of multiple resting-state networks \cite{SmithASMRNetwork}. Kovacevich examined comments to ASMR videos and found positive comments appreciated the calming or relaxing effects \cite{KovacevichASMRMusicRelaxing}. For social and intimate experiences, Klausen argued that ASMRtists leverage binaural sounds and haptic interactions to create a form of embodied presence and distant intimacy with the viewers \cite{KlausenSafeandSound}. Smith and Snider also suggested ASMR performers intentionally express feelings of intimacy and affection to the viewers \cite{SmithAffectIntimacy}. Recent HCI research explored the use of ASMR effects in wearable technologies for enchantment and slow experiences \cite{DevendorfASMRMundane}. Studies on food-eating videos (colloquially known as ``Mukbang'') found ASMR a key motivator for video watching \cite{WeberFoodInfluenceSocialMedia, TangMukbang}. YouTube study-with-me videos also use ASMR effects \cite{LeeStudyWithMe}. 
\par
RQ3 seeks to obtain an initial understanding of viewers' responses to different ASMR experiences through comment analysis. Prior work found that people can have different feelings with the same ASMR trigger \cite{PoerioInsomniaRelaxation, FredborgMindfulness} and don't publicly share ASMR experiences with others \cite{BarrattASMRFlowLike, TangMukbang}. Comments represent immediate and direct user reactions to a video and analyzing comments is a more straightforward way to capture viewers' feelings than rating by external participants \cite{KovacevichASMRMusicRelaxing}. Word analysis of YouTube comments is a common approach to infer the influence of videos on viewers \cite{SiersdorferYouTubeComments, AlhassanSelfHarmYouTube, ScottYouTubeNOAAComments}. In RQ3, we measure how ASMR viewers comment on three common feelings of ASMR identified in prior research: social connection and intimacy, sensory perception, and relaxation and sleepiness. Considering the difficulties of manually annotating a large number of comments and subjectively rating viewers' feelings, a mixed-method of Linguistic Inquiry and Word Count (LIWC) software \cite{Pennebaker2015TheLIWC2015} and pointwise mutual information (PMI) \cite{PMI} is used. LIWC has been scientifically validated to analyze people's social and emotional expression on social media \cite{ChoudhuryTwitterMentalHealth, FastEmpath}. PMI is also a lexicon-based method to identify topically important keywords \cite{ChoudhuryTwitterMentalHealth, RhoPoliticalHastag}. Both methods are widely used in prior HCI research to imply psychological processes from social media data (e.g., \cite{ChoudhuryTwitterMentalHealth, FastEmpath, RhoPoliticalHastag}). Then we compare the word frequencies in comments of videos with different interaction modalities and parasocial attractions. 

\section{ASMR Video Data}
We collected recent ASMR videos from active video creators to analyze ASMR videos on YouTube. The ASMR videos were crawled using the YouTube Data API\footnote{https://developers.google.com/youtube/v3} with the search seed ``ASMR.'' In the first step, we searched ASMR videos on Jun 11, 2020, and Oct 20, 2020, to collect a list of videos posted in the prior three months, respective to each search date. This step lets us identify active channels that were recently posting ASMR videos. Then the crawler collected all available videos belonging to those active channels to form a raw dataset. 227,133 videos were returned from YouTube. For each video, we requested the YouTube API to return 300 top-level comments\footnote{YouTube Data API does not specify how the returned comments are selected and ordered. https://developers.google.com/youtube/v3/docs/comments/list}. Since comments belonging to a video are analyzed together, and popular videos may have numerous comments, collecting up to 300 per video ensures all videos have a similar amount of comments. 
\par
Titles and tags are processed to filter out non-English videos. We exclude non-English videos due to difficulties in the data tagging and categorization. Videos without ``ASMR'' in the titles are removed since this work focuses on intentional ASMR videos with a clear ASMR theme and is designed for this experience (an ASMRtist may post non-ASMR videos). We exclude videos shorter than 5 minutes ($N=9676, 4.26\%$) due to many of them being previews of full videos and compilations of short video clips from multiple ASMR videos. We also only keep videos posted between Jan 1, 2020, and Jun 01, 2020, to ensure the videos reflect the latest creation styles and have enough time to receive comments. We remove videos with fewer than 50 comments (31.39\% of videos) to ensure videos had enough comments for word analysis. 
\par
After filtering, 85,734 videos are kept for data sampling. These videos come from 697 different channels. Then we randomly sample 200 videos for grounded theory analysis. We sample up to 10 videos per channel for the final data analysis. There are many channels with less than ten videos in our dataset -- the eventual sampling results in 2830 videos for data annotation. The data collection overlaps with the emergence of COVID-19, but through a rough examination, we didn't notice a significant mention of COVID-19 in the videos. The IRB (Institutional Review Board) office at the authors' institute has reviewed the entire research process and exempted this research from ethics board review (see Appendix \ref{dpia} for more information).

\section{Methods}

\subsection{Grounded Analysis}
Analysis in prior studies identified triggers from a small video sample or a few popular ASMR roleplays. Therefore, we choose to conduct a grounded theory analysis to extract common interaction modalities and parasocial attraction techniques. Grounded theory data analysis has been widely used to inductively derive models of social processes \cite{EavesGroundedTheory, Charmaz2015GroundedTheory}. This work follows open, axial, and selective coding procedures to generate and verify modality and attractiveness subcategories. We randomly sample 200 videos from 166 ASMRtists for the grounded analysis \cite{NiuASMRCSCW}. 
\par
\begin{table}[!ht]
    \caption{Subcategories and definition of the five multimodal interactions in ASMR videos}
    \centering
    \scalebox{0.7}{
        \begin{tabular}{|m{1em}|P{2.2cm}|p{9.2cm}|}
            \hline
            & Category & Definition\\
            \hline
            \multirow{7}{*}{\rotatebox[origin=c]{90}{Visual}} & Face-to-face & The ASMRtist looks at the camera to mimic face-to-face interactions with the viewer.\\
            \cline{2-3}
            & Mukbang & The ASMRtist presents and consumes large quantities of food (Mukbang)\\
            \cline{2-3}
            & Object only & The ASMRtist interacts with physical objects without showing their faces \\
            \cline{2-3}
            & Serve people & The ASMRtist performs a treatment/service on another person \\
            \cline{2-3}
            & Images & Static image(s) or black screen \\
            \cline{2-3}
            & Gaming & Video shows clip(s) of gaming, with or without the ASMRtist in view \\
            \cline{2-3}
            & Animals & The video has animals as the characters \\
            \hline
            \multirow{9}{*}{\rotatebox[origin=c]{90}{Sound}} & Object & Sounds made by interacting with a physical or liquid object by tapping, scratching, pouring, spraying, etc.\\
            \cline{2-3}
            & Whispering & Whispering or talking in a low volume \\
            \cline{2-3}
            & Mouth & Sounds made with mouth by eating, drinking, lip smacking, tongue clicking, kissing, licking, or sucking \\
            \cline{2-3}
            & Body\&cloth & Sounds made by touching/brushing/scratching themselves, another person, or a fake/silicon body in the video  \\
            \cline{2-3}
            & Ambience & Ambient and background sounds emitted from a real or fake environment \\
            \cline{2-3}
            & Mic & Sounds made by interacting with the microphone \\
            \hline
            \multirow{7}{*}{\rotatebox[origin=c]{90}{Touch}} & Viewer & The ASMRtist reaches to the viewer with their hands or tools in front of the camera \\
            \cline{2-3}
            & Objects & The ASMRtist clicks, taps, scratches, squeezes, or rubs physical objects  \\
            \cline{2-3}
            & Own body & The ASMRtist touches their own head, body, clothes by rubbing, scratching, combing, applying makeup, etc.  \\
            \cline{2-3}
            & Real person & The ASMRtist uses their hands or tools to interact with another real person in the video  \\
            \hline
             & Taste & The ASMRtist eats or drinks for more than half of the video  \\
            \hline
            \multirow{9}{*}{\rotatebox[origin=c]{90}{Scenario}} & Service & The video is a treatment or service roleplay in which the ASMRtist acts as a service provider and the viewer acts as a customer/patient (e.g., massage, haircut, makeup application, clinical exam, interview, customer service).   \\
            \cline{2-3}
            & Fantasy & The video is a roleplay in which the ASMRtist acts as a character in a fantasy, surreal, or otherwise unrealistic scenario (e.g., historical/anime/comics character)   \\
            \cline{2-3}
            & Romance & The video is a roleplay in which the ASMRtist acts as an intimate partner and directly interacts with the viewer intimately or romantically.   \\
            \hline
        \end{tabular}
    }
    \label{tab:codingbook_modality}
\end{table}

\begin{table}[!ht]
    \caption{Subcategories and definition of the three parasocial attractiveness in ASMR videos}
    \centering
    \scalebox{0.7}{
        \begin{tabular}{|P{1em}|P{2.5cm}|p{8.9cm}|}
            \hline
            & Category & Definition\\
            \hline
            \multirow{5}{1.5cm}{\rotatebox[origin=c]{90}{Social}} & Talk To  & The ASMRtist talks to or reads to the viewer \\
            \cline{2-3}
            & Talk with & The ASMRtist pretends to talk with or chat with the viewer, pretending the viewer responds to the ASMRtist \\
            \cline{2-3}
            & Gesture and text & The ASMRtist makes eye contact with the viewer and uses body language/closed captions/texts to communicate with the viewer  \\
            \hline
            \multirow{8}{1.5cm}{\rotatebox[origin=c]{90}{Proximity}} & Closeup & One of the 3 camera shot scales (Extreme closeup, Closeup, Medium closeup) in which ASMRtists placing themselves close to the camera\\
            \cline{2-3}
            & Medium & One of the 2 camera shot scales (Medium shot and Medium-full shot) in which ASMRtists placing themselves in medium distance to the camera \\
            \cline{2-3}
            & Fullshot & The ASMRtists show full body in the camera \\
            \cline{2-3}
            & No face & Static image(s), black screen, or no human face in the video  \\
            \cline{2-3}
            & Partial face  & Showing half-face (upper or lower half face)  \\
            \hline
            \multirow{7}{1.5cm}{\rotatebox[origin=c]{90}{Task}} & Treatment and service & The ASMRtist performs treatment/service on the viewer or another person in the video (e.g., massage, makeup application, interview, office visit, hypnosis, Reiki, etc.)  \\
            \cline{2-3}
            & Common activity & The ASMRtist engages in common daily activity(s) such as painting, writing, folding clothes, preparing food, or applying makeup to themselves.   \\
            \cline{2-3}
            & Eat and drink & The ASMRtist eats and/or drinks in the video   \\
            \hline
        \end{tabular}
    }
    \label{tab:codingbook_parasocial}
\end{table}

\begin{figure*}[ht]
    \centering
    \scalebox{1}{
       \includegraphics[width=\linewidth]{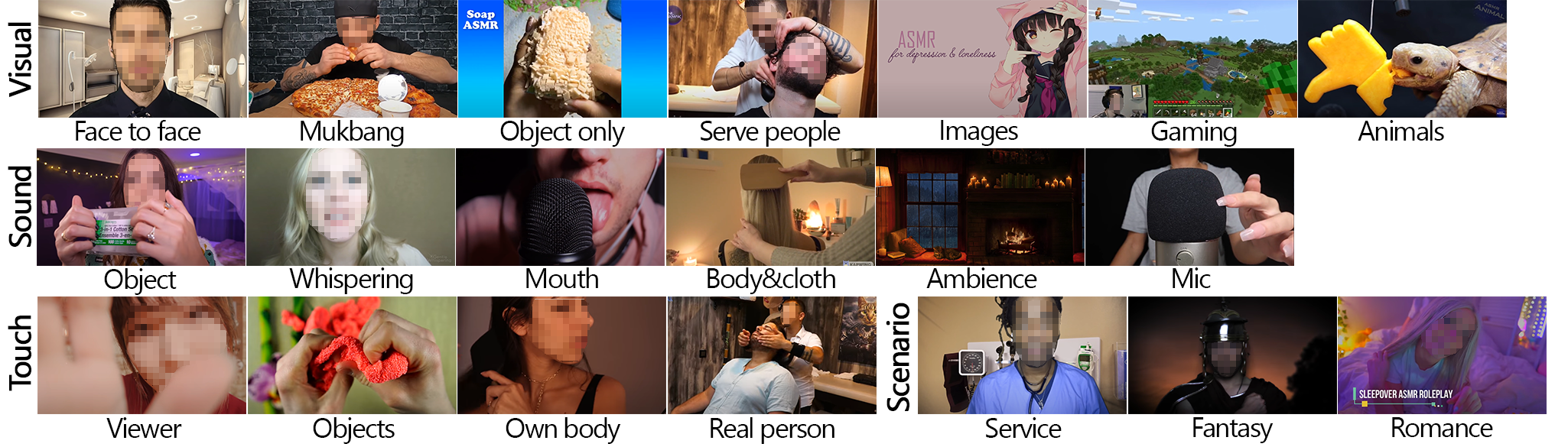}
    }
    \caption{Examples of visual, sound, touch, and scenario subcategories}
    \Description{Examples of visual, sound, touch, and scenario subcategories}
    \label{fig:modality_example}
\end{figure*}

In the open coding phase, two of the authors of this research watched 50 videos each and take notes on the visual, sound, touch, taste, and scenario triggers described in \cite{RichardBrainTingles}. The example multimodal interactions can be found in Figure \ref{fig:modality_example}. For parasocial attractiveness, the authors annotated how the ASMRtists simulate communication with their viewers (social attraction), where the ASMRtist is situated in proximity to the camera (spatial proximity), and the tasks the ASMRtists perform (task attraction). 
\par
\begin{figure}[ht]
    \centering
    \scalebox{1}{
       \includegraphics[width=\linewidth]{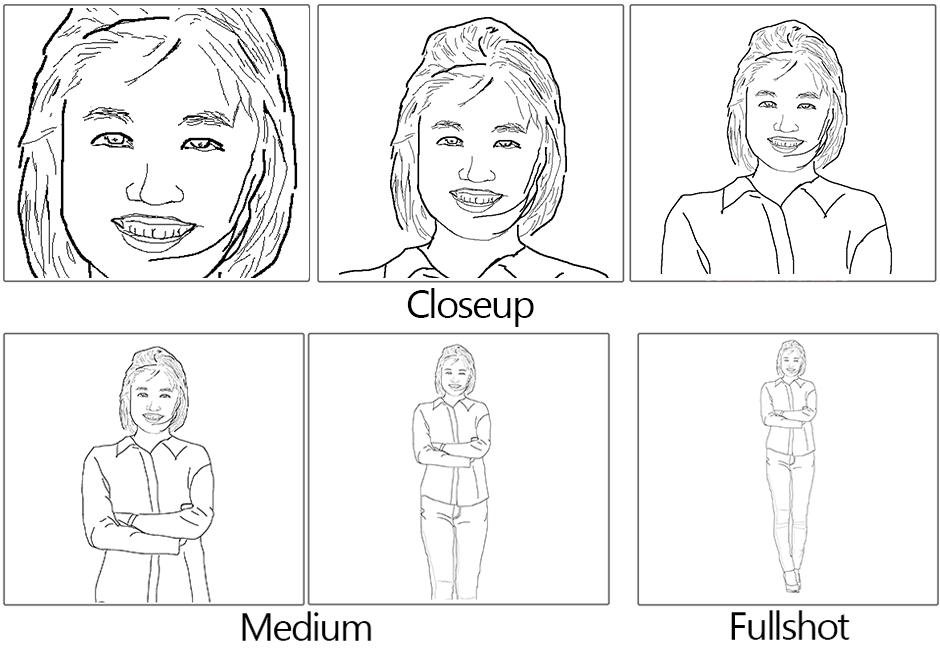}
    }
    \caption{Closeup, medium, and full shot examples in spatial proximity}
    \Description{Closeup shows chest or above, or as close as just show the performer's face. Medium is from knees or waist above. Fullshot shows the whole body.}
    \label{fig:shotsize}
\end{figure}

For axial coding, the two authors used the affinity diagramming approach to summarize these notes and develop subcategories of interaction modalities and parasocial attractions (see Table \ref{tab:codingbook_modality} and Table \ref{tab:codingbook_parasocial} for the codebook). The categorization of social attraction identifies that ASMRtists may communicate with the viewer in the form of a one-sided talk, or may chat with the viewers as in a back-and-forth conversation by pausing and waiting for the viewer to reply. Some other ASMRtists make ASMR videos without any human voice on the audio track, instead using gestures and text to communicate. For spatial proximity, we focus on categorizing the proximity with which ASMRtists positioned themselves in relation to the camera. We adopt the shot scales used in film and TV\footnote{https://www.studiobinder.com/blog/types-of-camera-shots-sizes-in-film/} (Figure \ref{fig:shotsize}) to annotate how the ASMRtist is displayed within the video frame (including \textit{extreme closeup}, \textit{closeup}, and \textit{medium closeup}), medium distance (including \textit{medium shot} and \textit{medium-full shot}), or showing the whole body from head to toe (\textit{full shot}). The annotation of the tasks performed by the ASMRtists finds three main categories of activities with clear goals. Treatment and service tasks seek to perform actions such as massage or haircut on the viewer. Some videos perform everyday tasks such as painting, writing, or applying makeup. Other videos demonstrate eating or drinking a large quantity of food (called Mukbang videos on YouTube \cite{TangMukbang}). 
\par

In selective coding, two authors annotated the remaining 100 videos using the codebook to validate the subcategories and obtain the inter-rater agreements between experts. Audio and touch were annotated as multi-categorical values. Visual, taste, scenario, social, physical, and task were annotated as single-categorical values. After annotation, 12 of 100 videos were removed due to unavailability (e.g., deleted, private, age-restricted, or non-English). Fleiss Kappa with Jaccard distance was used to calculate between expert agreement. For the 88 videos, all multimodal and parasocial categories reach substantial agreements with kappa scores between 0.62 and 0.88 (Table \ref{tab:fleiss_kappa}). Social and task have relatively lower agreement due to the differences in deciding if the ASMRtist talks to or talks with the viewer (e.g., a video is tagged differently because the ASMRtist mostly whispers by herself but also greets the viewer) and whether an activity is considered common and daily (e.g., one rater feels mixing makeup slime is a common activity while the other rater annotate it as a non-task video). Sound and touch have a lower agreement because they are multi-choice categories. Then the third author annotated disagreed answers independently to solve discrepancies and generate 88 expert annotations. The expert annotations were used to assess the accuracy of annotations completed on Amazon Mechanical Turk.

\begin{table}[!ht]
    \caption{Agreement scores between experts and between experts and MTurk participants. Calculated using Fleiss' Kappa with Jaccard method.}
    \centering
    \scalebox{0.64}{
        \begin{tabular}{|P{3cm}||P{1cm}|P{1.2cm}|P{0.9cm}||P{1cm}|P{1cm}|P{1cm}|P{1cm}|P{1cm}|P{1.3cm}|}
            \hline
            & Social & Proximity & Task & Visual & Sound & Touch & Taste & Scena.\\
            \hline
            Between\linebreak experts & 0.65 & 0.83 & 0.62 & 0.84 & 0.68 & 0.64 & 0.88 & 0.78 \\
            \hline
            Between experts and MTurk & 0.73 & 0.79 & 0.72 & 0.80 & 0.68 & 0.67 & 0.93 & 0.76\\
            \hline

        \end{tabular}
    }
    \label{tab:fleiss_kappa}
\end{table}

\subsection{Video Data Annotation}
This work uses Amazon Mechanical Turk (MTurk) to annotate the ASMR videos. Each task consists of two steps. Each participant was asked to watch each video for three minutes in the beginning, one minute in the middle, and one minute in the end. Then the participant was asked to annotate the five multimodal interactions and three parasocial attractions by answering multi-choice questions. Example pictures were provided to explain each visual and proximity subcategory. Example video clips were provided to explain each sound and touch subcategory. A qualification test was performed to pre-screen qualified participants. To be qualified, the MTurkers must indicate that they have watched at least 10 ASMR videos before, do not feel ASMR videos disturbing or unsatisfying, and experience a tingling sensation and relaxation after watching ASMR videos. To test participants' ASMR knowledge, a pre-screen question asked them to pick two typical ASMR videos from the other four non-ASMR videos. To ensure annotation quality, we only invited MTurkers who have completed more than 5000 tasks on MTurk with an approval rate greater than 97\%. A simple math question and a question asking participants to choose two ASMR videos from four video descriptions were deployed in each task as attention tests. MTurkers must answer both attention tests correctly to get the work accepted. Otherwise, the task is rejected and re-released to other participants.
\par
Before annotating all the data, we first test the agreement between MTurk workers and the expert annotations. MTurkers also completed the 88 videos that the experts have annotated. The annotations of all subcategories between experts and MTurkers reached a substantial agreement, with kappa scores ranging from 0.67 to 0.80 (Table \ref{tab:fleiss_kappa}). At the end, the annotation was completed by 47 MTurk participants with an average completion time of 8.9 minutes. Each accepted task was paid at the rate of USD \$1.50. MTurkers report 167 videos with problems (99 unavailable, 28 age-restricted, and 40 non-English). After removing the 167 videos, 2663 videos are used for final data analysis. These videos have an average of 225,143.69 views ($SD=654717.5$) and 5180.89 likes ($SD=9554.74$). 

\subsection{Comment Data Processing}
483,583 comments of the 2663 videos were collected for analysis, with each video having 181.59 comments on average ($SD=87.81$). Among 2663 videos, 687 videos have 300 comments. The comment analysis uses two different text processing methods, LIWC \cite{Pennebaker2015TheLIWC2015} and PMI \cite{PMI}, to obtain words related to social connection and intimacy, sensory perception, and relaxation and sleepiness. For each video, we merge all the collected comments into one corpus and use LIWC to calculate the percentages of words related to social processes and sensory perception. Social processes include words like ``mate,'' ``talk,'' and ``they'' and all non-first-person-singular personal pronouns, as well as verbs that suggest human interaction (talking, sharing) \cite{Pennebaker2015TheLIWC2015}. Sensory perception includes bodily and perceptual words. The body category under biological processes contains words such as ``cheek,'' ``hands,'' and ``spit.'' Perceptual processes recognize words related to perception, including ``look,'' ``heard,'' and ``feeling.'' We also generate a text document with all comments combined and obtained the emotional tone and social-, body-, and perception-word percentages in the entire comment corpus. This step allows us to examine the overall sentiment in ASMR video comments and compare ASMR comments with the base rates of expressive writing, natural speech, and Twitter data \cite{Pennebaker2015TheLIWC2015}.
\par
LIWC does not provide words related to intimacy, relaxation, and sleep processes. Therefore, we use the pointwise mutual information (PMI) technique \cite{PMI} to recognize words and phrases associated with those processes. Similar approaches have been widely applied in social media analysis (e.g., \cite{ChoudhuryTwitterMentalHealth, RhoPoliticalHastag}). The PMI measures the likelihood that two terms occur together in a corpus. The PMIs are calculated based on the 8,914,289 comments from the filtered 85,734 videos. Comments are pre-processed to remove stop words and punctuation to retain only meaningful words (including regular words and emojis). Words are then processed to find bigrams of common phrases. The keywords to generate associated word lists are ``intimate,'' ``relax,'' and ``sleep.'' To filter out too-rare and too-common words, identified associated words must appear in at least 1000 comments and no more than 1/10 of all comments. We choose the top 3\% of the qualified terms with the highest PMIs as the word lists for each keyword. Table \ref{tab:pmi_words} lists example words associated with each keyword. Then we apply a similar approach as in LIWC to count the percentages of words from each list in each video's comments.
    
\begin{table}[!ht]
    \caption{Example words associated with intimacy, relaxation, and sleep identified by PMI. The word list is used to count the percentages of words associated with each feeling.}
    \centering
    \scalebox{0.7}{
        \begin{tabular}{|M{1.3cm}|M{1.1cm}|p{9.3cm}|}
            \hline
            Keyword & Number of words & Example words with top PMIs \\
            \hline
            \multirow{4}{*}{intimate} & \multirow{4}{*}{85} & intimate, intimacy, connection, sexual, relationships, sensual, romantic, personal, desire, events, atmosphere, relationship, emotionally, casual, decision, approach, scenario, witness, creation, client, private, audience, sacrifice, strangely, destiny, partner, remain, alternate, detail, interact\\
            \hline
            \multirow{4}{*}{relax} & \multirow{4}{*}{157} & relax, informative, unwind, tense, stressful day, wonderfully, educational, incredibly, stressful, entertain, stevie, strangely, lavender, atmosphere, extremely, meds, super, amazingly, movements, peaceful, manner, drift, stress, visually, surprisingly, serum, ambient, traditional\\
            \hline
            \multirow{4}{*}{sleep} & \multirow{4}{*}{155} & sleep, have trouble, peacefully, drift, schedule, clinic, nights, meds, pills, autoplay, 4am, insomnia, alarm, pill, tonight, trouble, brush teeth, aid, auto play, medication, 1am, night, induce, nightmares, wake, 2am, help, put, pm, 3am\\
            \hline
        \end{tabular}
    }
    \label{tab:pmi_words}
\end{table}

\subsection{Statistical Method}
The visual, taste, scenario annotations, and the three parasocial subcategories are stored as multi-categorical nominal variables. Sound and touch, the two subcategories with multiple possible choices, are saved as dummy variables (1 is containing the interaction, otherwise 0). For RQ2, Pearson's Chi-squared test (contingency table) is used to identify significant associations between multimodal interaction and parasocial attraction subcategories. For comment analysis in RQ3, we first perform regression analysis to identify multimodal and parasocial subcategories that significantly predict each feeling word percentage. For each feeling word, two least-squared regression (LSR) models are built, with one using multimodal interactions as independent variables and the other using parasocial attraction variables. The two models are multivariate regressions with all modality factors (or all attraction factors) serving as independent variables simultaneously. We perform posthoc analysis with the Steel-Dwass method for each significant factor to identify differences between subcategory pairs. Nonparametric comparisons are performed due to the word frequencies being not normally distributed. In all statistical testings, the significant threshold ($alpha=0.05$) is adjusted with the Bonferroni method.

\section{Results}
\subsection{RQ1: Multimodal Interactions in ASMR Videos}
\begin{figure*}[!ht]
    \centering
    \scalebox{1}{
    \begin{tabular}{ll}
    \begin{minipage}{0.55\textwidth}
        \centering
        \includegraphics[width=\linewidth]{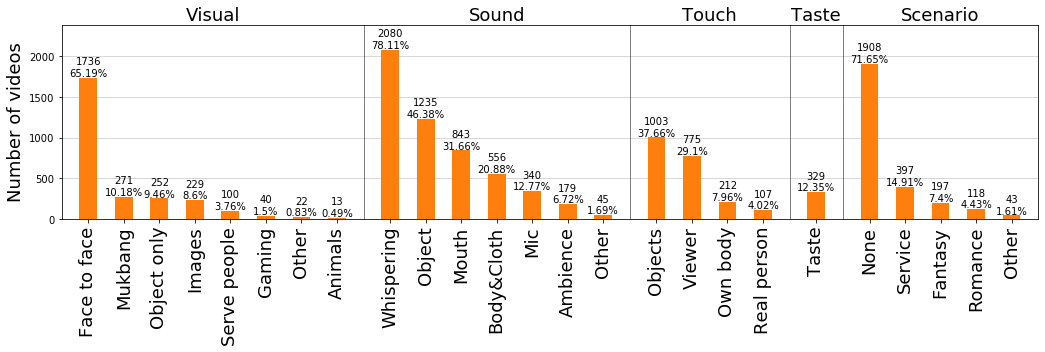}
         \Description[]{Face to face, whispering, touching objects, and none scenario have the highest number of videos}
    \end{minipage}%
    \begin{minipage}{0.45\textwidth}
        \centering
        \includegraphics[width=\linewidth]{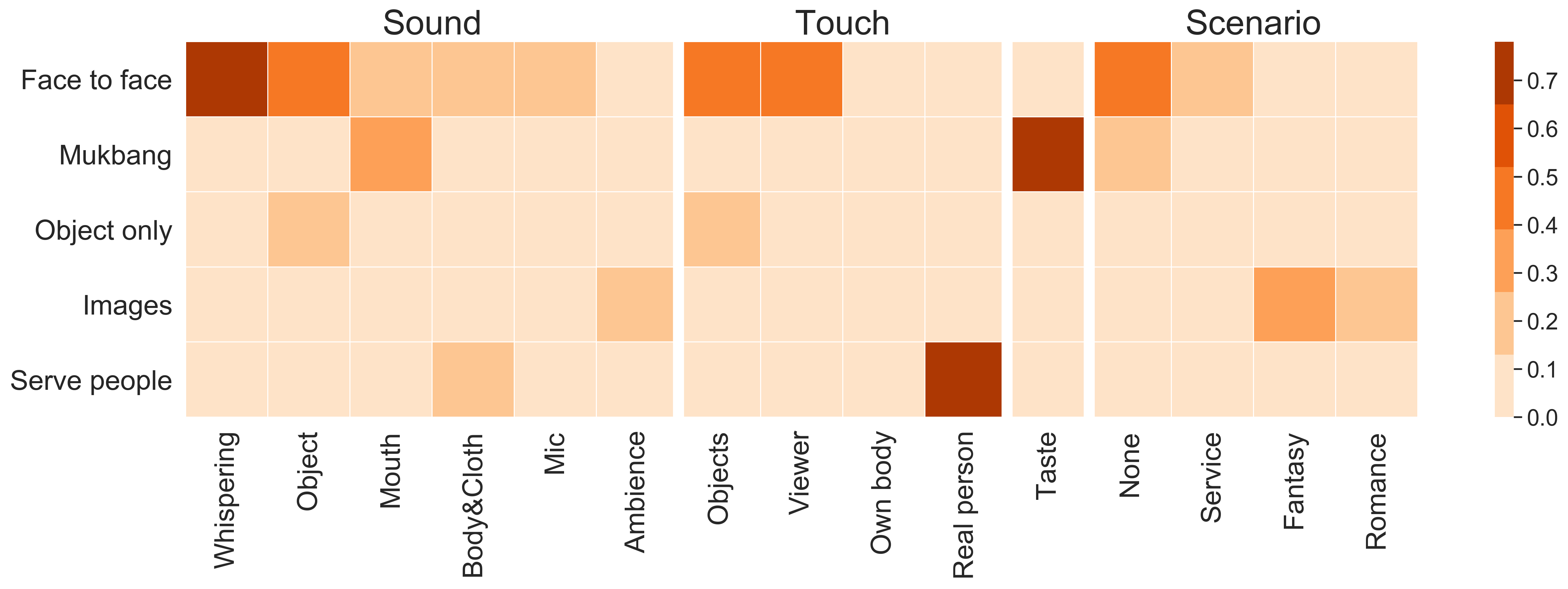}
         \Description[]{The overlaps of visual sub-categories between visual interactions and sound, touch, taste and scenario.}
    \end{minipage}
    \end{tabular}
    }
\caption{Left: The distribution of videos in the subcategories of the five interaction modalities. Right: The co-appearance rates of visual triggers and triggers of other modalities, calculated using the Jaccard index.}
\Description{Face-to-face is the highest in visual subcategories. Whispering is the highest in sound subcategories. Objects are highest in touch subcategories. And none is the highest in scenario subcategories. Face-to-face has more co-appeared interactions.}
\label{fig:multimodal}
\end{figure*}

RQ1 seeks to overview the interaction modalities used in ASMR videos and suggest how prevalent triggers are (see Figure \ref{fig:multimodal} left). \textbf{Visual}. Face-to-face is the most common visual setting. Around 2/3 of videos present ASMRtists themselves in front of the camera. The ASMRtists also perform triggers in other modalities like making whispering and object sounds, manipulating objects, or reaching to the camera at the same time (see Figure \ref{fig:multimodal} right). Mukbang, object only, and images are also common visual settings. Only a small proportion of ASMR videos use video games, animals, or other visual interactions. \textbf{Sound}. We notice ASMRtists tend to mix multiple sound effects in ASMR videos. Only 896 (33.65\%) videos contain only one type of sound. 1081 (40.59\%) of videos use two audio triggers. 687 (25.80\%) videos use three or more different types of sound effects. The most common sound in ASMR videos is whispering and soft speaking. Other common sound effects include object sounds, mouth effects, body and cloth sounds, mic effects, and ambient sounds. \textbf{Touch}. More than half ($N=1573, 59.07\%$) of ASMR videos use at least one touch trigger in the video. The most common touch interaction is touching objects in the video to generate tingling sounds. 29.1\% of videos have ASMRtists reaching toward the camera and pretended to touch the viewers' face or body. Less than 10\% of videos contain touching ASMRtists' own body parts or a person in the video. \textbf{Taste}. Tasting is not a commonly used interaction modality in ASMR videos. Only 12.35\% of videos use tasting triggers, mostly in Mukbang videos (Figure \ref{fig:multimodal} right). \textbf{Scenario}. 71.65\% of videos do not use any roleplays in the videos. The most common roleplay is in service scenarios in which ASMRtists perform services or treatment processes. Less than 10\% of videos are fantasy or romance roleplays. 
\par
The analysis of the distribution of multimodal interactions in ASMR videos suggests most ASMRtists choose to look at the camera to mimic face-to-face interactions with the viewer. ASMR videos are sound-diverse and rich. The most common sound interactions are whispering to the viewers and sounds made by manipulating trigger objects. Around 1/3 of ASMR videos use touch interaction by touching objects and/or touching the viewers. Taste interaction is used to perform Mukbang videos. The majority of the ASMR videos do not have roleplays and plotted scenarios. The most common ASMR scenario is service or treatment roleplays.


\subsection{RQ2: Parasocial Attractions and ASMR Experience Patterns}
\begin{figure*}[!ht]
    \centering
    \scalebox{1}{
    \begin{tabular}{ll}
    \begin{minipage}{0.55\textwidth}
        \centering
        \includegraphics[width=\linewidth]{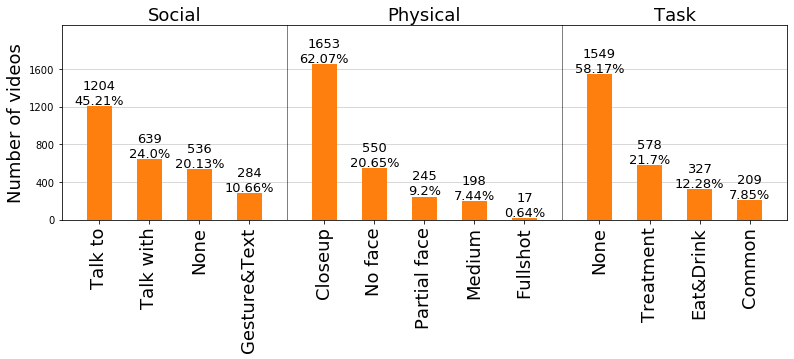}
         \Description[]{Talk-to, closeup, and no-task videos have the highest number of videos.}
    \end{minipage}%
    \begin{minipage}{0.45\textwidth}
        \centering
        \includegraphics[width=\linewidth]{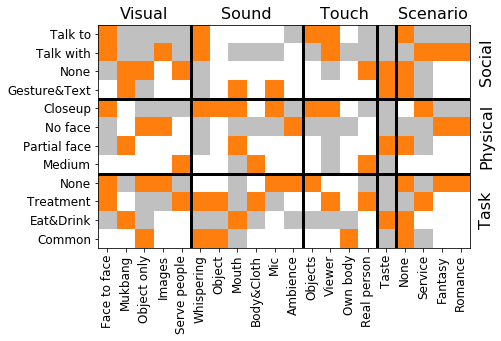}
         \Description[]{The significant associations between interaction modalities and parasocial attractions.}
    \end{minipage}
    \end{tabular}
    }
\caption{Left: the distribution of videos in subcategories of parasocial attractions. Right: The significant associations between subcategories of multimodal interactions and parasocial attractions. Orange squares are significant positive associations. Grey squares are significant negative associations.}
\Description{Talk-to is the highest in social subcategories. Closeup is the highest in proximity subcategories. None is the highest in task subcategories.}
\label{fig:parasocial}
\end{figure*}

RQ2 examines parasocial attractions in ASMR videos and their associated interaction modalities to evoke ASMR experiences. Figure \ref{fig:parasocial} left illustrates their distribution, and Figure \ref{fig:parasocial} right shows all the positive and negative associations.
\par

\subsubsection{ASMR for Social Experiences}
\begin{figure}[!ht]
\centering
\scalebox{1}{
    \begin{tabular}{ll}
    
        \begin{subfigure}{.49\linewidth}
        \includegraphics[width=\linewidth]{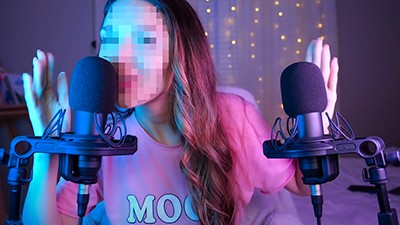}
        \caption{(a) The ASMRtist softly whispers to the viewer.\newline}
        \end{subfigure}
        & 
        
        \begin{subfigure}{.49\linewidth}
        \includegraphics[width=\linewidth]{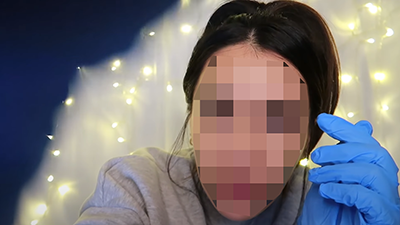}
        \caption{(b) The ASMRtist talks with viewer and pretends to reach viewer's face in a treatment.}
        \end{subfigure}
        \\
        
        \begin{subfigure}{.49\linewidth}
        \includegraphics[width=\linewidth]{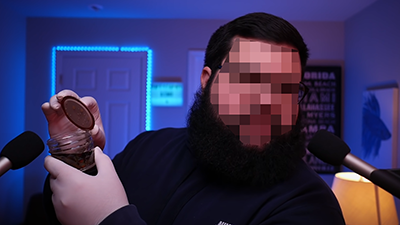}
        \caption{(c) The ASMRtist manipulate a bottle near the mic while introducing it in whispers.}
        \end{subfigure}
        
        &
        \begin{subfigure}{.49\linewidth}
        \includegraphics[width=\linewidth]{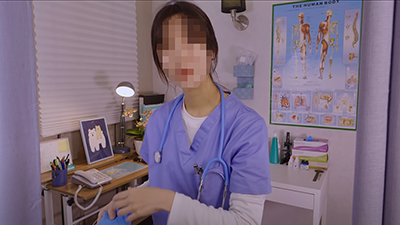}
        \caption{(d) The ASMRtist performs a conversation in a nerve exam on the viewer in a treatment roleplay.}
        \end{subfigure}
        \\ 
        \begin{subfigure}{.49\linewidth}
        \includegraphics[width=\linewidth]{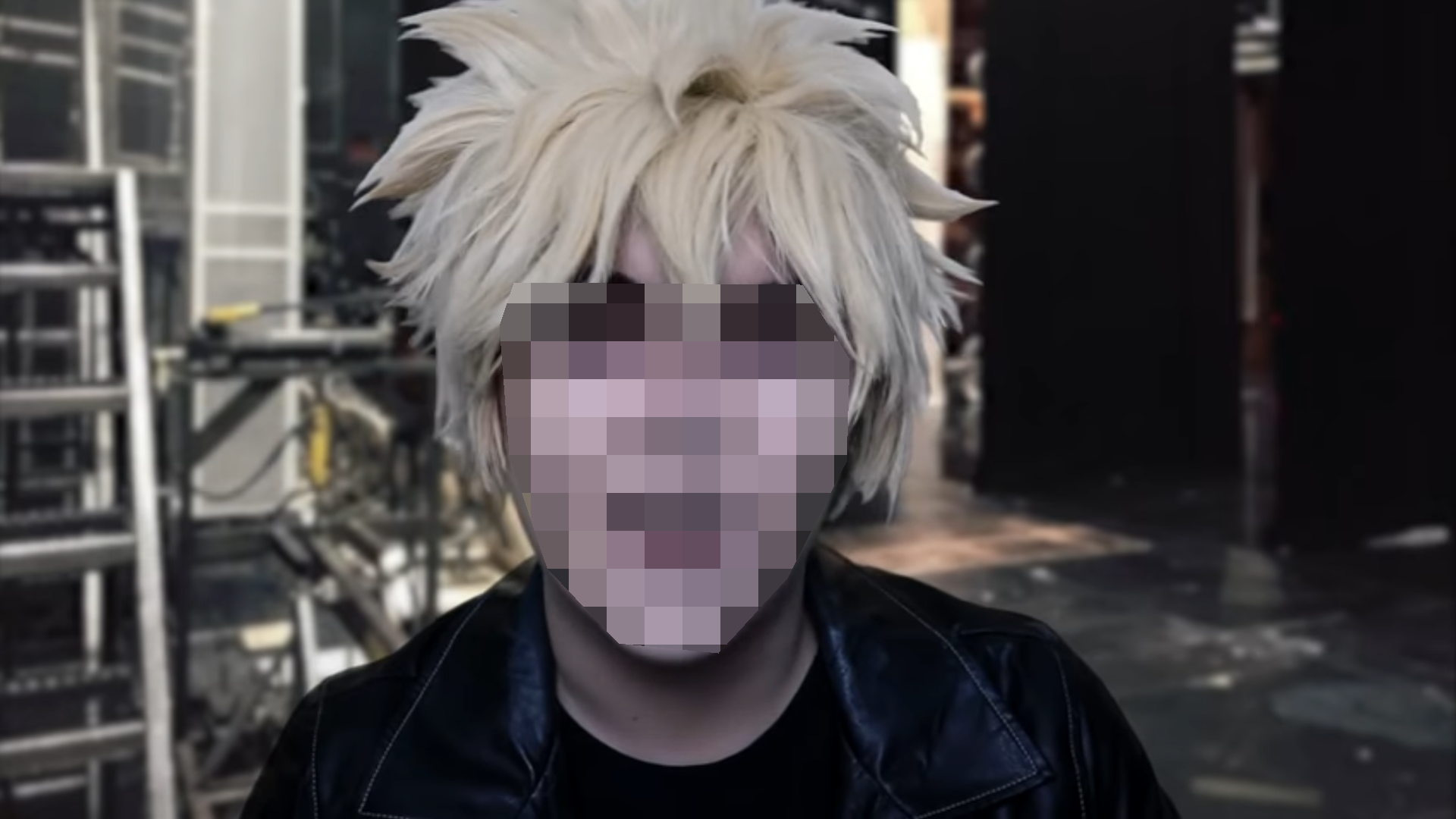}
        \caption{(e) The ASMRtist pretends to recruit the viewer into their punk band in a fantasy roleplay.}
        \end{subfigure}
        & 
        \begin{subfigure}{.49\linewidth}
        \includegraphics[width=\linewidth]{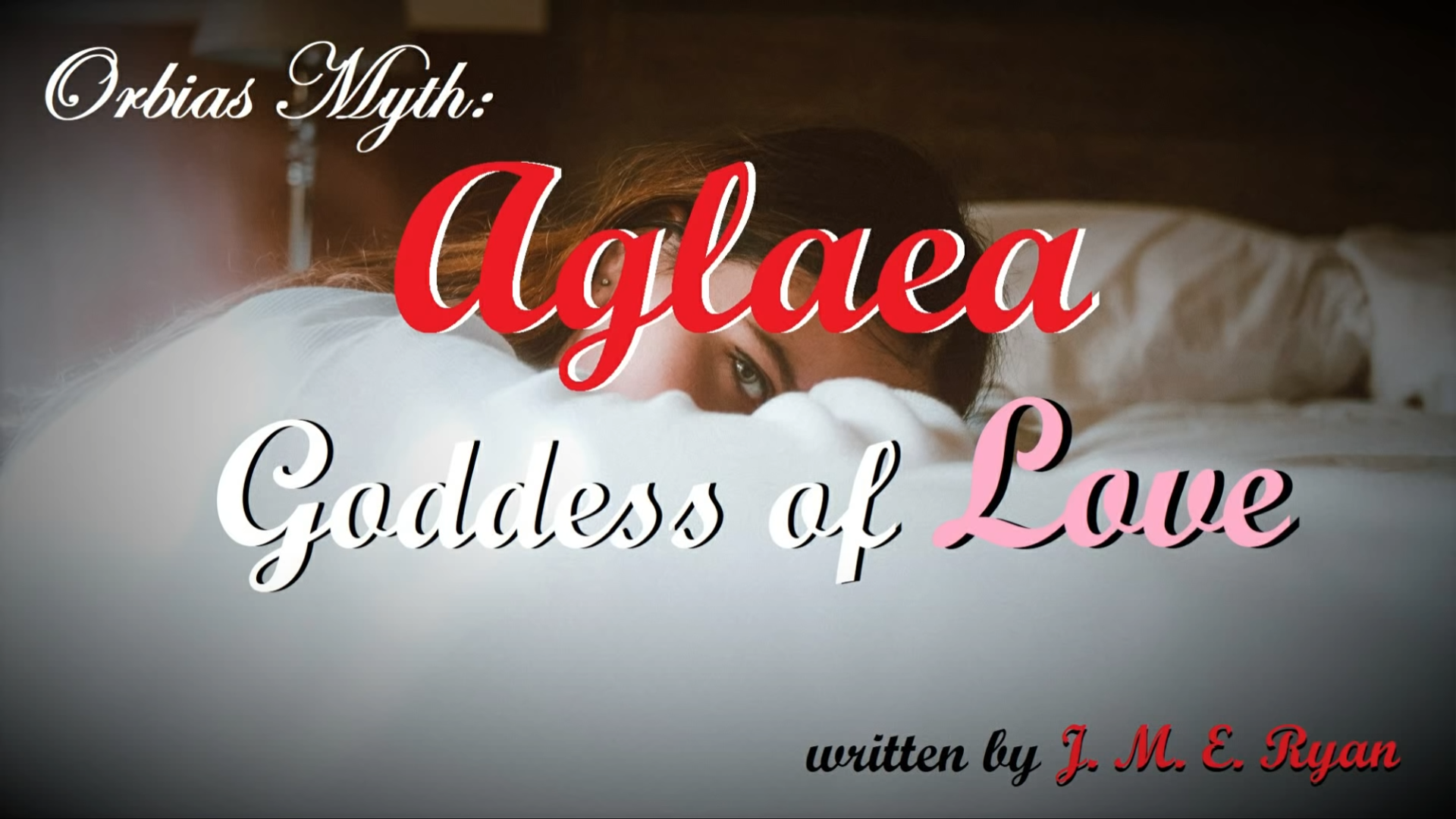}
        \caption{(f) The ASMRtist talks with the viewer as the Goddess of Love in a romantic, image-only roleplay.}
        \end{subfigure}
    \end{tabular}
}
\caption{Example ASMR videos showing ASMR interactions for social attraction.}
\Description{Example ASMR videos to explain social connections.}
\label{fig:socialexample}
\end{figure}

The analysis of social attraction shows that 1843 (69.21\%) of the videos have the ASMRtists talking to or talking with the viewer (Figure \ref{fig:parasocial} left). Pearson's chi-squared test suggests the face-to-face presentation ($visual.face\_to\_face$), whispering sounds ($sound.whispering$), and virtually touching the viewer through camera-reaching ($touch.viewer$) are significantly associated with talk-to or and talk-with videos ($social.talk\_to$ and $social.talk\_with$). Videos that talk to the viewers are also significantly associated with touching objects ($touch.objects$). Talking with the viewers is also associated with all three types of roleplays (Figure \ref{fig:parasocial} right). Among all 1843 talk-to and talk-with videos, 1366 have the ASMRtists interacting with the viewers in ``face-to-face'' settings and making whispering sounds (e.g., Figure \ref{fig:socialexample}-a). 645 videos also pretend to touch the viewers through camera reaching (e.g., Figure \ref{fig:socialexample}-b). 504 videos talk to the viewers while the ASMRtists touch objects to make tingling sounds (e.g., Figure \ref{fig:socialexample}-c). Talk-with videos use scenarios in which the ASMRtist roleplays a service provider ($N=273$, Figure \ref{fig:socialexample}-d), a fantasy character ($N=108$, Figure \ref{fig:socialexample}-e), or an intimate partner ($N=81$, Figure \ref{fig:socialexample}-f). Although most talk-with videos show the performer looking at the viewer face-to-face, 116 out of 639 talk-with videos are videos with static or no images ($visual.images$, e.g., Figure \ref{fig:socialexample}-f). Non-socializing videos are associated with Mukbang ($visual.mukbang$), object-only ($visual.object\_only$), serving other people ($visual.serve\_people$), touching another person in the video ($touch.real\_person$), tasting ($taste$), and non-roleplays ($scenario.none$). 
\par

The high percentage of videos with conversational content suggests that social connection is a common experience incorporated by ASMRtists. ASMR performers deliver social connection experiences with multimodal interactions such as face-to-face whispering, hand-reaching, and one-sided or back-and-forth conversation. The ASMRtists tend to touch objects to generate tingling sounds while talking to the viewers. ASMRtists also perform service, fantasy, and romantic roleplays to engage the viewer in an emulated conversation, in which many ASMRtists pretend that they can hear the viewer's responses. Videos without socialization are videos of food eating, serving another person, or merely manipulating objects.

\subsubsection{ASMR for Intimate Interaction}
\begin{figure}[!ht]
\centering
\scalebox{1}{
    \begin{tabular}{ll}
        \begin{subfigure}{.49\linewidth}
        \includegraphics[width=\linewidth]{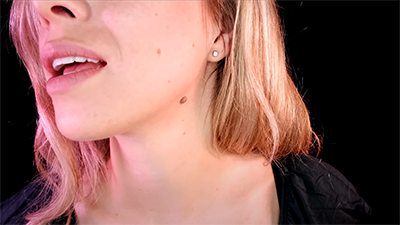}
        \caption{(a) An ASMRtist whispers near the camera/mic to simulate close-ear speaking.}
        \end{subfigure}
        &
        \begin{subfigure}{.49\linewidth}
        \includegraphics[width=\linewidth]{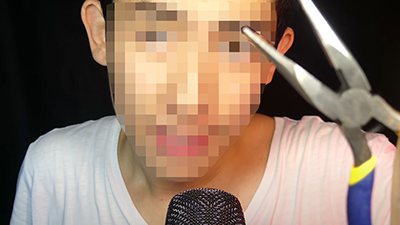}
        \caption{(b) The ASMRtist manipulates pliers near the camera to generate tingling sound.}
        \end{subfigure}
        \\
        \begin{subfigure}{.49\linewidth}
        \includegraphics[width=\linewidth]{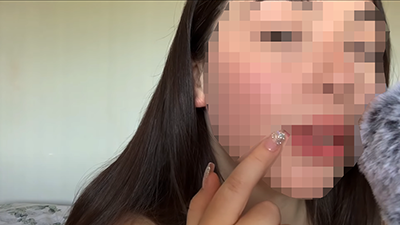}
        \caption{(c) The ASMRtist smacks lips near the mic to emulate close-ear feelings.}
        \end{subfigure}
        &
        \begin{subfigure}{.49\linewidth}
        \includegraphics[width=\linewidth]{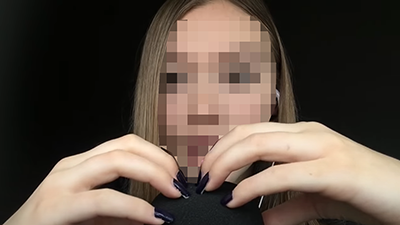}
        \caption{(d) The ASMRtist scratches the mic and whispers trigger words (coconut, pickle, etc.).}
        \end{subfigure}
        \\
        \begin{subfigure}{.49\linewidth}
        \includegraphics[width=\linewidth]{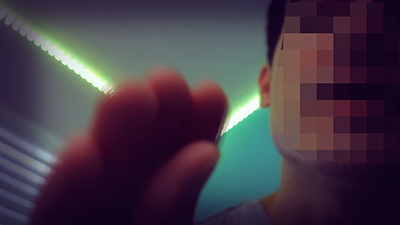}
        \caption{(e) The ASMRtist pretends to do makeup on viewer's face while the viewer is asleep.}
        \end{subfigure}
        & 
        \begin{subfigure}{.49\linewidth}
        \includegraphics[width=\linewidth]{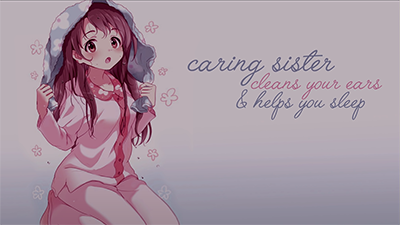}
        \caption{(f) An audio-only ASMR, where ASMRtist emulates cleaning viewer's ears as a sister.}
        \end{subfigure}
    \end{tabular}
}
\caption{Example ASMR videos showing ASMR interactions for intimacy.}
\Description{Example ASMR videos to explain physical intimacy.}
\label{fig:physicalexample}
\end{figure}

The result of spatial proximity suggests that the majority (1868, 70.15\%) of the ASMR videos have the ASMRtists presenting their full faces in closeup, medium, and full shot camera distances (Figure \ref{fig:parasocial} left). Only 550 (20.65\%) videos do not have human appearances, and 245 (9.2\%) videos show partial faces. Closeup is the most used shot scale used in ASMR videos, suggesting most ASMRtists seek to simulate nearness with the viewer by positioning themselves in close camera proximity. The association analysis suggests that face-to-face ($visual.face\_to\_face$), whispering sounds ($sound.whispering$), object sounds ($sound.object$), mouth sounds ($sound.mouth$), mic effects ($sound.mic$), and the touching of objects ($touch.objects$) and viewers ($touch.viewer$) significantly associate with closeup camera proximity ($proximity.closeup$, Figure \ref{fig:parasocial} right). In all 1653 closeup videos, 1490 videos whispers to the viewer near the camera (e.g., Figure \ref{fig:physicalexample}-a). 809 videos have ASMRtists making various trigger sounds (e.g., Figure \ref{fig:physicalexample}-b). 591 and 245 closeup videos contain mouth effects (e.g., Figure \ref{fig:physicalexample}-c) and mic effects (e.g., Figure \ref{fig:physicalexample}-d). For touch interactions, closeup videos also consist of manipulating objects (e.g., Figure \ref{fig:physicalexample}-b) and pretending to touch the viewers (e.g., Figure \ref{fig:physicalexample}-e). In these videos, ASMRtists make mouth sounds near the mic or interact with the mic to engender the sound of physical closeness. Although some videos do not show the ASMRtist on screen, 59 of them simulate intimate and romantic roleplays in their conversations with the viewers (Figure \ref{fig:physicalexample}-f). 
\par
These results suggest that ASMRtists seek to use cameras and microphones to emulate intimate interactions with the viewers. Common approaches include positioning near the camera, near-mic whispering, manipulating objects, reaching hands to the camera, and making mouth and mic sounds. Even in videos without the performer's physical presence, ASMRtists can emulate intimate roleplays and conversations to express intimacy.

\subsubsection{ASMR for Activity Observation}
\begin{figure}[!ht]
\centering
\scalebox{1}{
    \begin{tabular}{ll}
        
        \begin{subfigure}{.49\linewidth}
        \includegraphics[width=\linewidth]{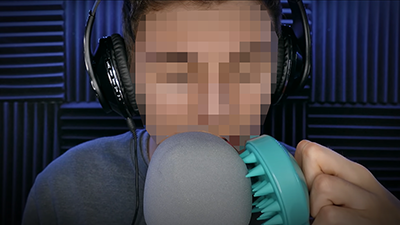}
        \caption{(a) The ASMRtist scratches the mic with a brush and makes random mouth sound.}
        \end{subfigure}
        & 
        \begin{subfigure}{.49\linewidth}
        \includegraphics[width=\linewidth]{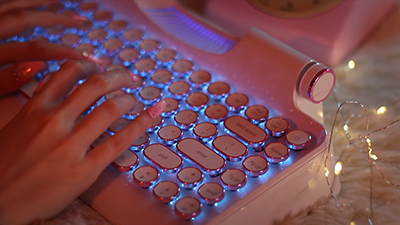}
        \caption{(b) An ASMRtist constantly typing on a keyboard for 1.6 hours without speaking.}
        \end{subfigure}
        \\ 
        \begin{subfigure}{.49\linewidth}
        \includegraphics[width=\linewidth]{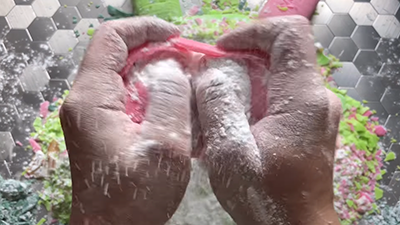}
        \caption{(c) An ASMRtist collapses soaps and starch without showing themselves or speaking.}
        \end{subfigure}
        &
        \begin{subfigure}{.49\linewidth}
        \includegraphics[width=\linewidth]{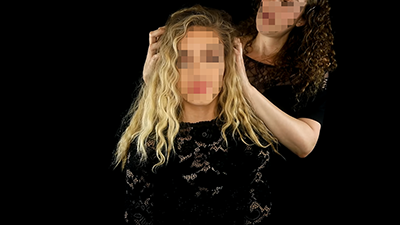}
        \caption{(d) The ASMRtist performs head massage on another person in the video.}
        \end{subfigure}
        \\
        
        \begin{subfigure}{.49\linewidth}
        \includegraphics[width=\linewidth]{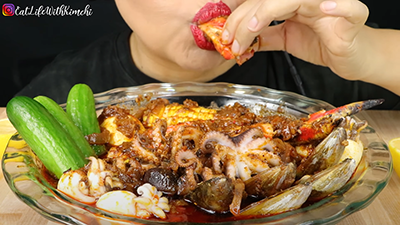}
        \caption{(e) The ASMRtist consumes a large quantity of food.}
        \end{subfigure}
        &
        \begin{subfigure}{.49\linewidth}
        \includegraphics[width=\linewidth]{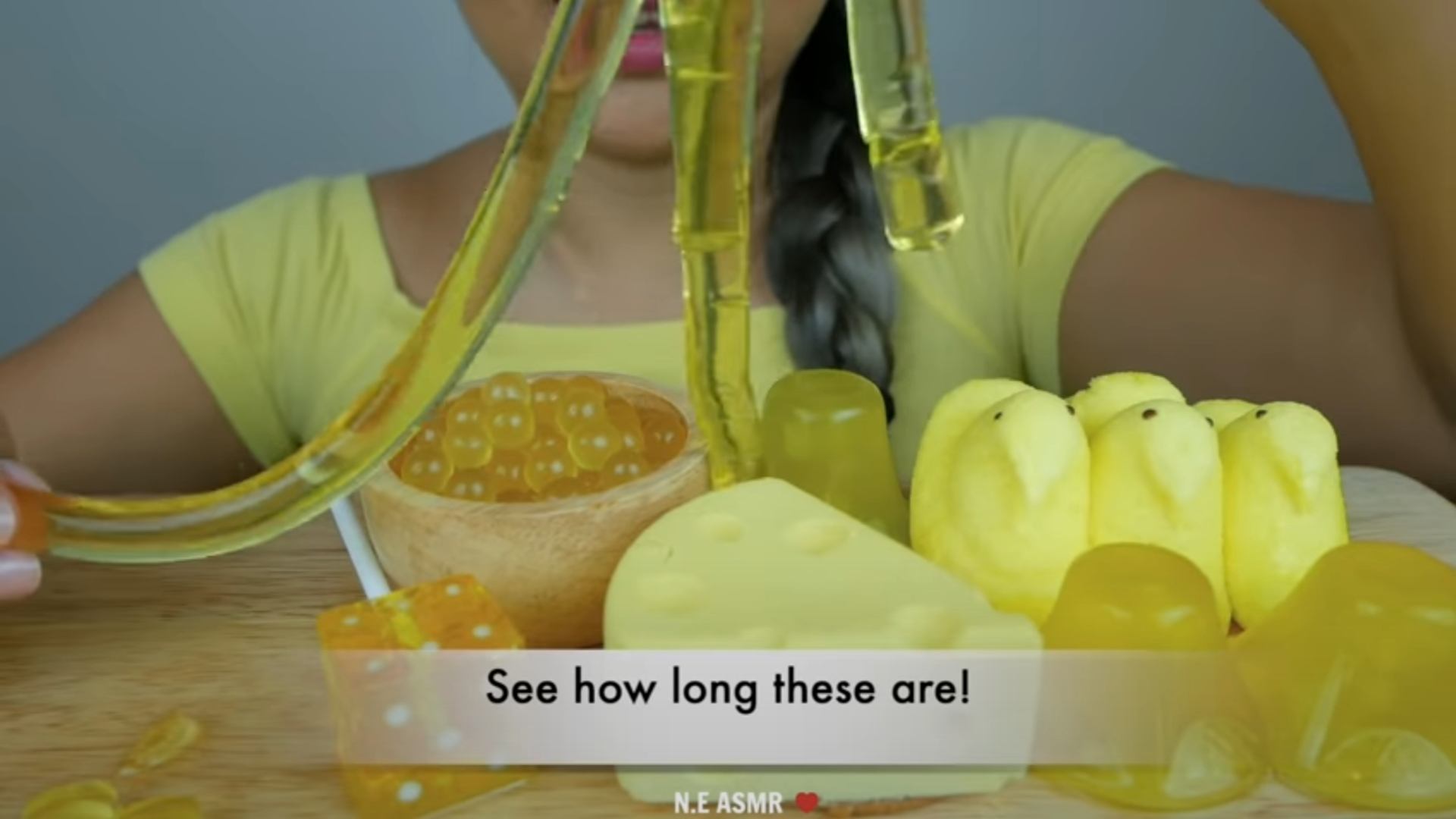}
        \caption{(f) The ASMRtist eats cakes and jelly and explains with captions.}
        \end{subfigure}
        \\
        \begin{subfigure}{.49\linewidth}
        \includegraphics[width=\linewidth]{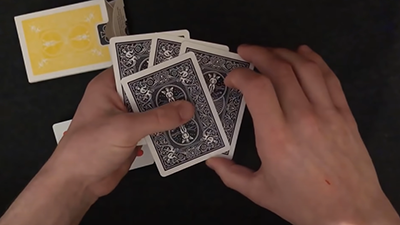}
        \caption{(g) The ASMRtist shows magic tricks with cards.}
        \end{subfigure}
        &
        \begin{subfigure}{.49\linewidth}
        \includegraphics[width=\linewidth]{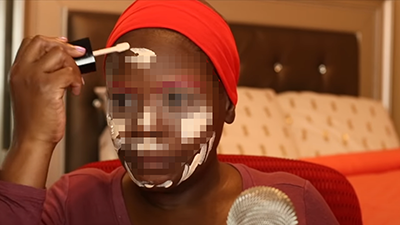}
        \caption{(h) The ASMRtist put makeup on face while chewing gum.}
        \end{subfigure}
    \end{tabular}
}
\caption{Example ASMR videos showing no-task and with-task ASMR interactions.}
\Description{Example ASMR videos to explain task and non-task activities.}
\label{fig:taskexample}
\end{figure}

The analysis of task attraction shows most of the videos do not contain a clear task. More than half of the videos ($N=1549, 58.17\%$) are $task.none$. The associations suggest three main types of videos that do not have specific tasks. The first type is ASMRtists using the face-to-face camera setting ($N=1085$) and chatting with the viewer and/or making random sounds (e.g., Figure \ref{fig:taskexample}-a). The second main type is chatting in fantasy or romance roleplays ($N=210$, e.g., Figure \ref{fig:socialexample}-f and \ref{fig:physicalexample}-f). The third type is object-only videos ($N=191$), in which ASMRtists manipulate trigger objects without meaningful purposes (e.g., Figure \ref{fig:taskexample}-b and c). More than 40\% of videos contain treatment, eating and drinking, and common daily tasks. Among 578 treatment videos, ASMRtists whisper to the viewer ($N=507$), make object ($N=317$) and body/cloth sounds ($N=164$), and emulate service scenarios ($N=381$). Treatment videos ($task.treatment$) are significantly associated with face-to-face ($visual.face\_to\_face$, $N=456$) and touching the viewer ($touch.viewer$, $N=304$), as those treatment videos pretend to perform the service on the viewer (e.g., Figure \ref{fig:socialexample}-b and d and Figure \ref{fig:physicalexample}-e). Treatment videos ($task.treatment$) are also significantly associated with performing service on another person ($visual.serve\_people$, $N=85$) and touching them ($touch.real\_person$, $N=83$, e.g., Figure \ref{fig:taskexample}-d). 337 videos consist of tasks of eating and drinking (e.g., Figure \ref{fig:taskexample}-e and f), most of which present a large quantity of food in the video ($visual.mukbang$, $N=259$) and make mouth sounds ($sound.mouth$, $N=301$). Mukbang task is significantly associated with non-social ($social.none$) and partial face presentations ($proximity.partial\_face$). The third major task is common daily activities (e.g., Figure \ref{fig:taskexample}-g and h). 216 (8.11\%) videos show daily tasks in which ASMRtists perform everyday activities. Common daily tasks ($Task.common$) are significantly associated with object-only presentations ($visual.object\_only$, $N=43$), whispering sounds ($sound.whispering$, $N=184$), and object sounds ($sound.object$, $N=122$). Daily tasks ($task.common$) are also significantly associated with touching own body ($touch.own\_body$, $N=38$) because these videos contain activities such as applying makeup (e.g., Figure \ref{fig:taskexample}-h). 
\par

These results indicate that ASMRtists tend to present activities without a clear, purposeful task. The taskless videos include mundane, repetitive, and unintentional actions, facing the viewers or only showing the trigger objects. Videos with particular tasks involve treating the viewer or another person, eating a large quantity of food, or other common everyday activities. Those activities are also considered soft, clicking, slow, or repetitive, which are likely to induce ASMR experiences \cite{BarrattASMRFlowLike}.
\par

\subsection{RQ3: Viewers' Comments about the Feelings of ASMR Experience}
\begin{figure}[!ht]
    \centering
    \scalebox{1}{
       \includegraphics[width=\linewidth]{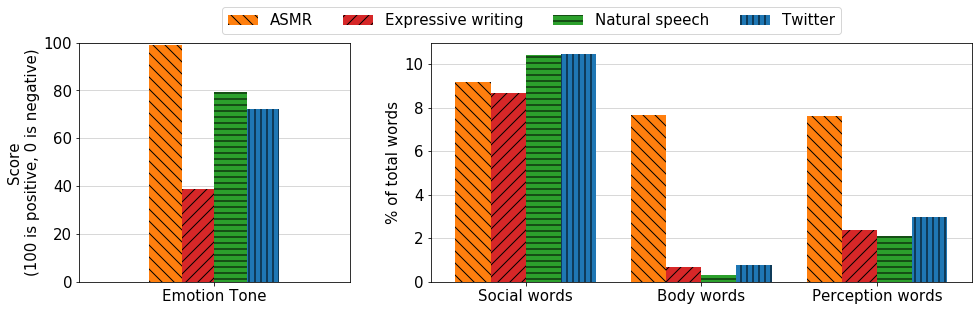}
    }
    \caption{Comparison of emotion tone and percentages of social, body, and perception words between ASMR videos and expressive writing, natural speech, and Twitter \cite{Pennebaker2015TheLIWC2015}}
    \Description{ASMR videos have a higher emotion tone, similar social words, more body and perception words than expressive writing, natural speech, and Twitter texts.}
    \label{fig:comment_overview}
\end{figure}

RQ3 explores viewers' feelings about different ASMR interactions by calculating the percentages of the LIWC and PMI identified keywords. We first compare the linguistic attributes of all ASMR video comments with the base rates of expressive writing, natural speech, and Twitter data \cite{Pennebaker2015TheLIWC2015}. Figure \ref{fig:comment_overview} shows the results. The emotion tone score of all ASMR comments is 99 (50 is neutral), higher than the other three types of textual data, indicating viewers' overall positive reaction to the ASMR videos. ASMR comments have comparable social word frequencies, suggesting viewers have similar social expression as in other texts. The body words (7.68\%) and perception words (7.63\%) in ASMR comments are higher than the other three texts. This shows that viewers write more in the comments about things associated with body and perception processes. The overall positive emotion and high frequency of body and perception words imply that viewers obtained sensational pleasure from watching the ASMR videos.

\begin{table*}[!ht]
    \caption{Results of multivariate LSR models that predict the percentages of feeling words in comments. Only the p-values of significant predictors are presented. Adjusted $\alpha=0.0083$. *Multi-choice factors, $p$ is from the dummy variable with the smallest p-value.}
    \centering
    \scalebox{0.8}{
        \begin{tabular}{|P{2.5cm}||P{1.2cm}|P{1.2cm}|P{1.2cm}|P{1.2cm}|P{1.2cm}|P{1.2cm}||P{1.2cm}|P{1.2cm}|P{1.2cm}|P{1.2cm}|P{1.2cm}|P{1.2cm}|P{1.2cm}|P{1.2cm}|}
            \hline
            \multirow{2}{*}{Dependent variable}  & \multicolumn{6}{c||}{LSR of Parasocial Attractions} & \multicolumn{8}{c|}{LSR of Multimodal Interactions} \\
            \cline{2-15}
             \rule{0pt}{8pt} & $F$ & $p$ & $r^2$ & $p_{social}$ & $p_{physical}$ & $p_{task}$ & $F$ & $p$ & $r^2$ & $p_{visual}$ & $p_{sound}$* & $p_{touch}$* & $p_{taste}$ & $p_{scenario}$ \\
            \hline
            Social words & 32.35 & <0.0001 & 0.11 & <0.0001 & <0.0001 & - & 18.44 & <0.0001 & 0.13 & <0.0001 & <0.0001 & - & - & - \\
            \hline
            Intimacy words & 24.03 & <0.0001 & 0.08 & <0.0001 & 0.0012 & <0.0001 & 16.92 & <0.0001 & 0.12 & 0.0016 & <0.0001 & - & - & 0.0001 \\
            \hline\hline
            Body words & 49.36 & <0.0001 & 0.16 & <0.0001 & <0.0001 & <0.0001 & 39.03 & <0.0001 & 0.25 & <0.0001 & <0.0001 & 0.0002 & 0.0060
            & 0.0003 \\
            \hline
            Perception words & 29.55 & <0.0001 & 0.10 & <0.0001 & <0.0001 & 0.0006 & 27.21 & <0.0001 & 0.18 & <0.0001 & <0.0001 & <0.0001 & - & <0.0001 \\
            \hline\hline
            Relax words & 43.28 & <0.0001 & 0.14 & 0.0003 & <0.0001 & <0.0001 & 30.42 & <0.0001 & 0.20 & <0.0001 & <0.0001 & <0.0001 & 0.00035 & <0.0001 \\
            \hline
            Sleep words & 36.73 & <0.0001 & 0.12 & 0.0065 & 0.0004 & <0.0001 & 23.03 & <0.0001 & 0.16 & - & <0.0001 & <0.0001 & <0.0001 & <0.0001 \\
            \hline

        \end{tabular}
    }
    \label{tab:comment_significance}
\end{table*}

\subsubsection{Social and Intimacy}
\begin{figure*}[!ht]
    \centering
    \scalebox{0.8}{
    \begin{tabular}{l}
    \begin{minipage}{\textwidth}
        \centering
        \includegraphics[width=\textwidth]{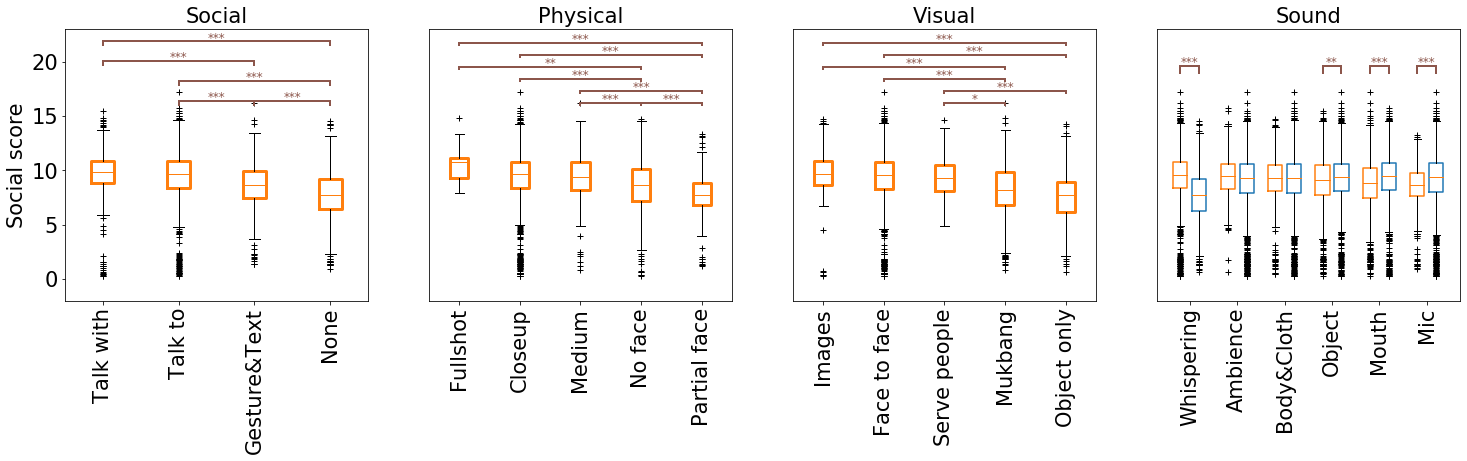}
         \Description[]{Talk-with, closeup, images, and whispering videos have more social words in comments.}
    \end{minipage}%
    \\
    \begin{minipage}{\textwidth}
        \centering
        \includegraphics[width=\linewidth]{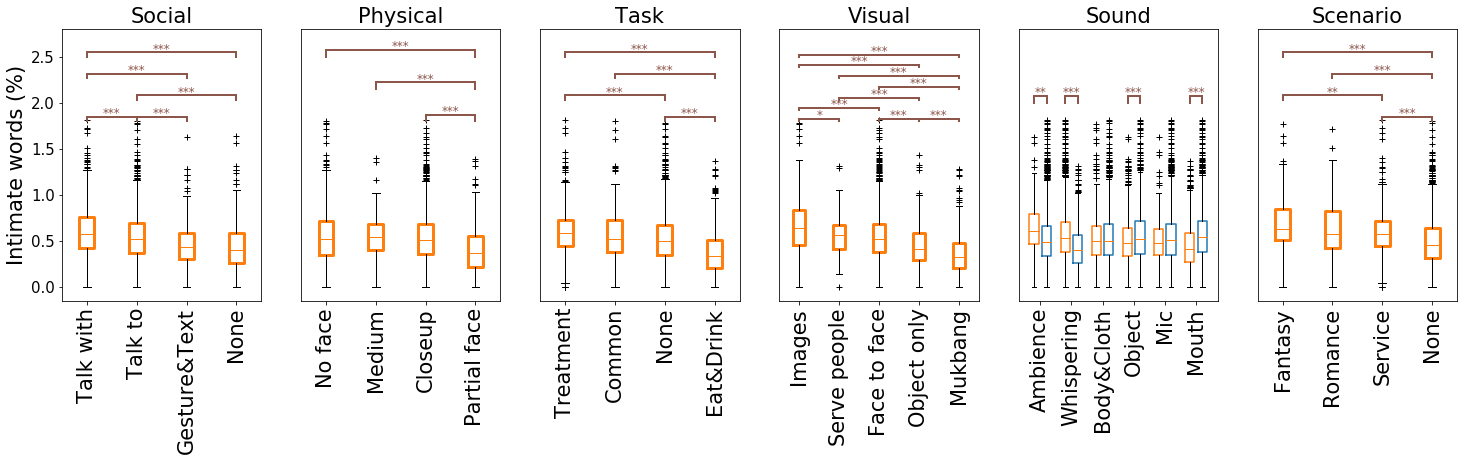}
         \Description[]{Talk-with, no face, treatment, images, ambience sound, and fantasy videos have more intimate words in comments.}
    \end{minipage}
    \end{tabular}
    }
\caption{Social and intimate word frequencies between subcategories of significant multimodal and parasocial predictors. Ordered by the average percentage in descending order. Horizontal bars show significant differences ($p$* < 0.05, $p$** < 0.01, $p$*** < 0.001).}
\Description{The highest social words are in talk with, closeup, images, and whispering videos. The highest intimate words are in talk with, no face, treatment, images, ambience sound, and fantasy videos.}
\label{fig:comment_social_intimacy}
\end{figure*}

The LSR model that predicts social word frequencies by the parasocial attraction subcategories suggests that parasocial attraction factors are significant predictors (Table \ref{tab:comment_significance}). The model that predicts social word frequencies by interaction modalities shows that visual and sound are significant predictors. The posthoc analysis shows that ASMR videos that leverage social attraction techniques lead to higher use of social words in the comments (Figure \ref{fig:comment_social_intimacy} top). The social word frequencies in talk-to and talk-with ($social.talk\_to$ and $social.talk\_with$) videos are significantly higher than gesture/text videos ($social.gesture\&text$) and non-social videos ($social.none$). Similarly, videos with ASMRtists whispering sounds have more social comments than videos without communication. For spatial proximity, videos with the ASMRtists being closeup ($proximity.closeup$), and medium distance ($proximity.medium$) have higher social word frequencies than videos without ASMRtists' appearances. With regard to visual modalities, videos that use visual settings of static images (mostly audio-only roleplays), face-to-face interaction, and serving people have higher social word frequencies than Mukbang and object manipulation videos. These results indicate that the social attraction techniques used in ASMR videos, such as talking to/with the viewers, showing themselves face-to-face, and whispering led viewers to express more social processes in the comments than non-social ASMR videos. 
\par

The LSR models which predict the intimate word frequency show that social, proximity, and task, are significant parasocial predictors (Table \ref{tab:comment_significance}). Visual, sound, and scenario are the significant multimodal interaction predictors. The posthoc analysis shows that talking with or to viewers leads to more comments related to intimacy. Figure \ref{fig:comment_social_intimacy} bottom shows the results. Videos with whispering sounds also have significantly more intimate words in the comments than videos without whispering. Mukbang videos have significantly lower intimate words in comments. The comparison of intimate words shows that comments to roleplay videos have more viewers' intimate expressions than videos without roleplays. Since many fantasy and romance roleplays are voice-only, videos without performers' faces ($proximity.no\_face$), with static images ($visual.images$), and with ambient sounds ($sound.ambience$) have significantly higher numbers of intimate words in the comments. 
\par
The comparison of social and intimate words suggests that the ASMR videos with social interactions -- such as presenting the ASMRtist in the videos and whispering to the viewers -- are more likely to receive viewers' social responses than ASMR videos without socialization. Roleplays lead to more intimate reactions in the comments than non-roleplay videos. In contrast, Mukbang and object-only videos have lower social and intimate expressions.

\subsubsection{Body and Perception}
\begin{figure*}[!ht]
    \centering
    \scalebox{1}{
    \begin{tabular}{l}
    \begin{minipage}{\textwidth}
        \centering
        \includegraphics[width=\linewidth]{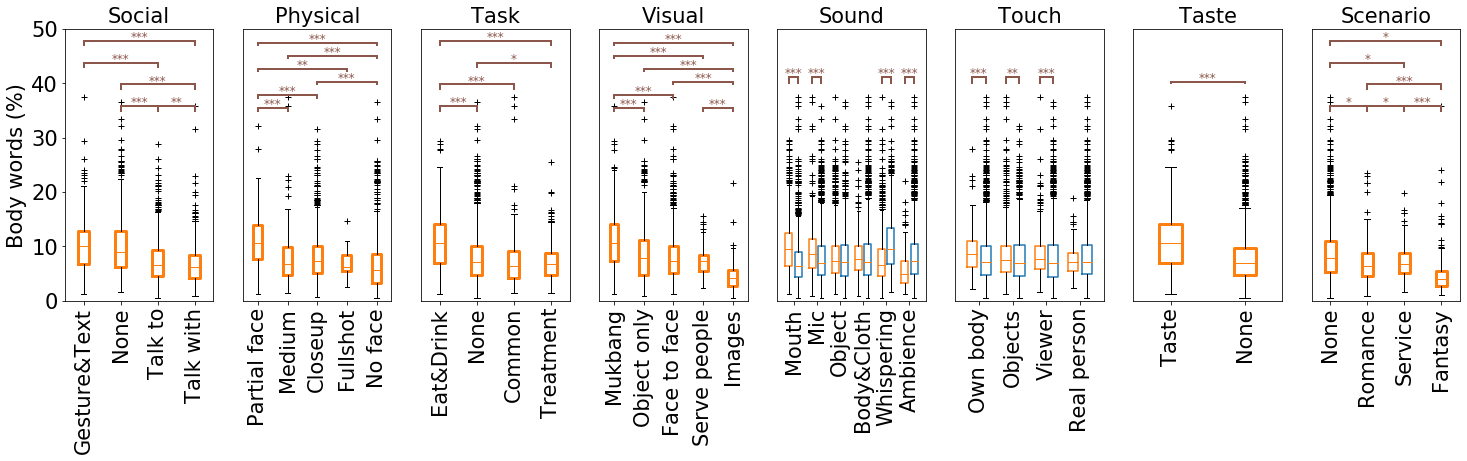}
         \Description[]{Gesture\&Text, partial face, eat\&drink, Mukbang, mouth sound, touching own body, and none-scenario videos have more body words in comments.}
    \end{minipage}%
    \\
    \begin{minipage}{\textwidth}
        \centering
        \includegraphics[width=\linewidth]{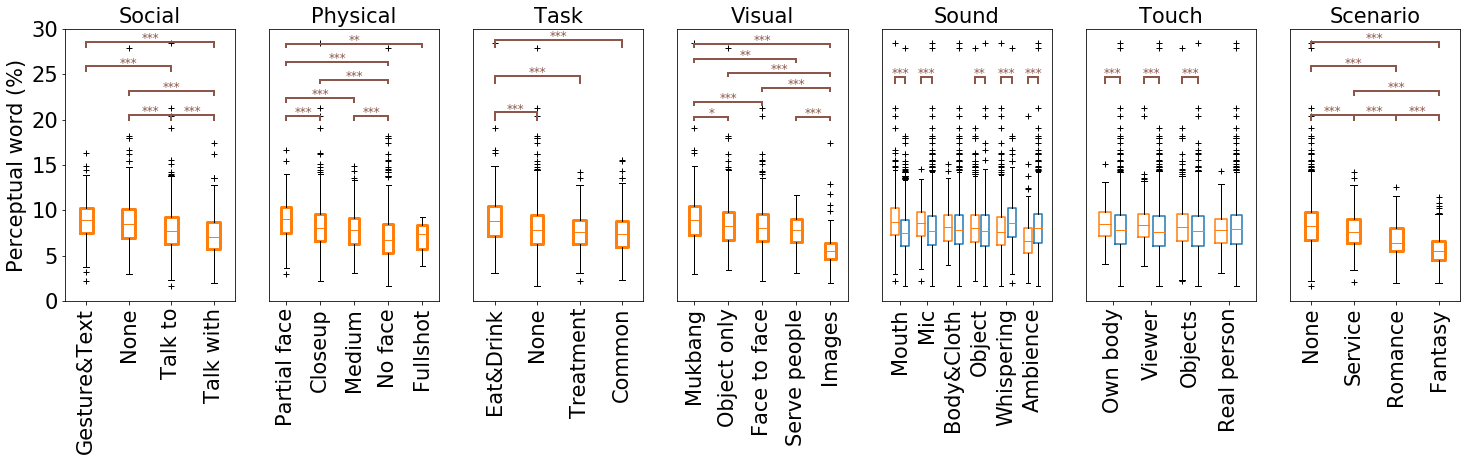}
         \Description[]{Gesture\&Text, partial face, eat\&drink, Mukbang, mouth sound, touching own body, and none-scenario videos have more perception words in comments.}
    \end{minipage}
    \end{tabular}
    }
\caption{Body and perception word frequencies between subcategories of significant multimodal and parasocial predictors. Ordered by the average percentage in descending order. Horizontal bars show significant differences ($p$* < 0.05, $p$** < 0.01, $p$*** < 0.001).}

\Description{The highest body words are in gesture and text, partial face, eat and drink, Mukbang, mouth sound, touch own body, taste, and none scenario videos. The highest perception words are in gesture and text, partial face, eat and drink, Mukbang, mouth, own body, and none scenario videos.}
\label{fig:comment_body_perception}
\end{figure*}

Viewers share their body and perceptual feelings or comment on ASMRtists' body or actions. The LSR model suggests social, spatial, and task attractions can significantly predict the use of body and perception words (Table \ref{tab:comment_significance}). All multimodal interaction subcategories are significant predictors of body words. Visual, sound, touch, and scenario are significant predictors of perception words. Posthoc shows that Mukbang videos lead to the highest sensory words in comments (Figure \ref{fig:comment_body_perception}). Eating or drinking videos ($task.eat\&drink$) have significantly more sensory responses than other task-oriented and taskless videos. Comments to videos that contain mouth sounds have significantly more body and perception words. Taste interactions lead to significantly more body words. Since Mukbang videos tend to show partial faces and only use gestures and text to communicate (e.g., Figure \ref{fig:taskexample}-e and f), the gesture/text ($social.gesture\&text$) and the partial face ($proximity.partial\_face$) videos have the highest sensory word use in the comments. We also noticed that videos without tasks ($task.none$) have higher body words. Non-roleplay videos ($scenario.none$) have the highest usages of body and perception words. Object-only videos ($visual.object\_only$, e.g., Figure \ref{fig:taskexample}-b and c) also have a higher mentioning of the body and perception words. Videos with object sounds ($sound.object$, e.g., Figure \ref{fig:socialexample}-c and Figure \ref{fig:physicalexample}-b) and mic sounds ($sound.mic$, e.g., Figure \ref{fig:physicalexample}-d and \ref{fig:taskexample}-a) have significantly more comments with body and perception words. These results imply that presenting and consuming a large quantity of food in ASMR videos, as well as videos without tasks or roleplays are more likely to induce viewers' feelings related to sensory perception. 
\par

\subsubsection{Relaxation and Sleepiness}
\begin{figure*}[!ht]
    \centering
    \scalebox{1}{
    \begin{tabular}{l}
    \begin{minipage}{\textwidth}
        \centering
        \includegraphics[width=\linewidth]{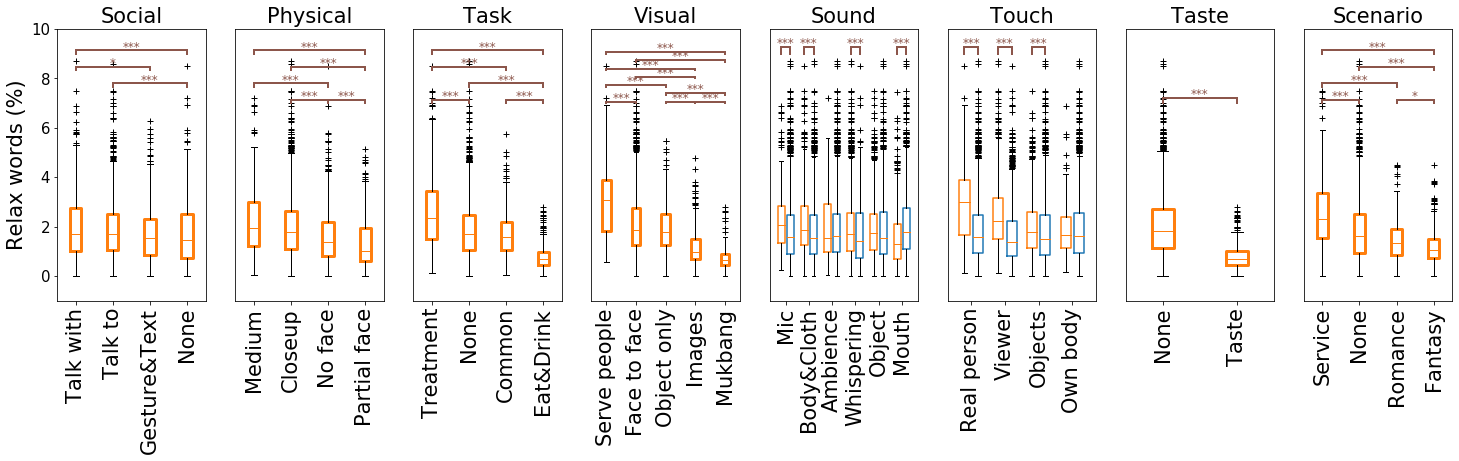}
         \Description[]{Talk-with, medium proximity, treatment task, serving people visual setting, mic sound, touching real person, none taste, and service scenario videos have more relax words in comments.}
    \end{minipage}%
    \\
    \begin{minipage}{\textwidth}
        \centering
        \includegraphics[width=\linewidth]{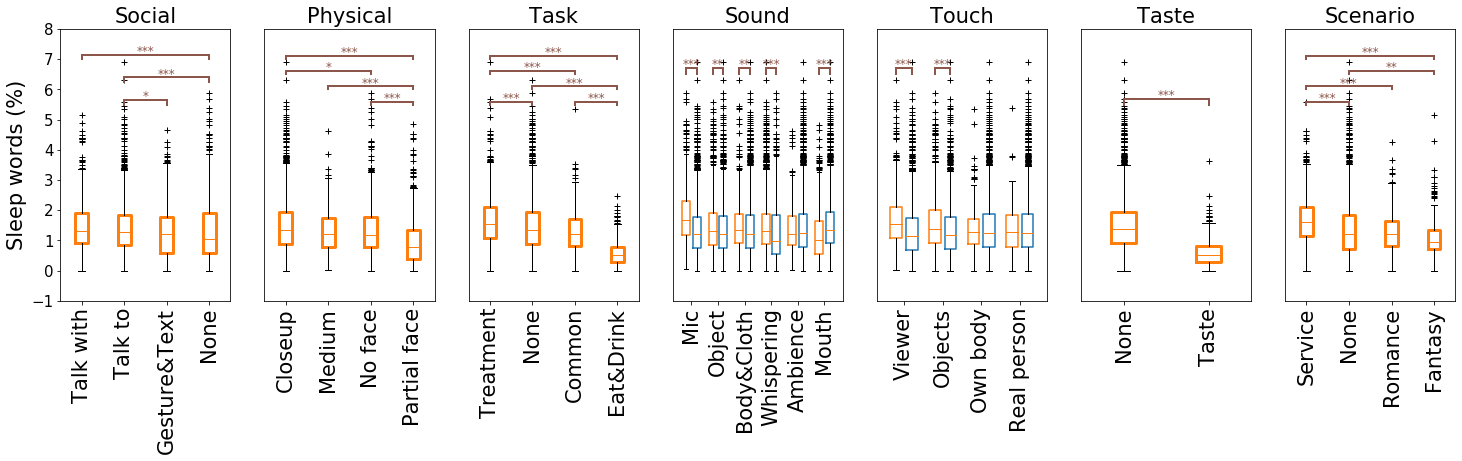}
         \Description[]{Talk-with, closeup proximity, treatment task, mic sound, touching the viewer, none taste, and service scenario videos have more sleep words in comments.}
    \end{minipage}
    \end{tabular}
    }
\caption{Relax and sleep word frequencies between subcategories of significant multimodal and parasocial predictors. Ordered by the average percentage in descending order. Horizontal bars show significant differences ($p$* < 0.05, $p$** < 0.01, $p$*** < 0.001).}
\Description{The highest relax words are in talk with, medium, treatment, serve people, mic sound, touch real person, none taste, and service videos. The highest sleep words are talk with, closeup, treatment, mic sound, viewer touch, none taste, and service videos.}
\label{fig:comment_relax_sleep}
\end{figure*}

The LSR model that predicts relaxation words shows that all parasocial and multimodal subcategories are significant predictors (Table \ref{tab:comment_significance}). The model that predicts sleep words suggests that all parasocial and multimodal subcategories except for visual interaction are significant predictors. Posthoc analysis between different categories shows that videos related to treatment and intimate interactions have the highest percentage of relax and sleep words in comments (Figure \ref{fig:comment_relax_sleep}). Videos with treatment performance ($task.treatment$ and $scenario.service$) have the highest percentages for both measurements. Videos that involve performing services on another person ($visual.serve\_people$ and $touch.real\_people$) also lead to more relaxation expression. Videos that pretend to touch the viewer by reaching toward the camera ($touch.viewer$) have significantly more sleep words than videos without this interaction. ASMR videos with mic sounds ($sound.mic$), in which ASMRtists get close up to the camera and make near-ear mic sound effects (e.g., Figure \ref{fig:physicalexample}-d and Figure \ref{fig:taskexample}-a), lead viewers to comment more about relaxation and sleepiness.
\par
These results imply that videos showing treatment processes and physical intimacy induce feelings of relaxation and sleepiness for viewers more often. Videos with near-ear microphone effects also incite feelings of relaxation and sleepiness. It should be noted that although the visual and touch settings of treatment ASMRs seek to simulate physical intimacy with the viewers, viewers do not express more intimacy in the comments. Instead, the emulation of close proximity interactions in treatments lets viewers express more relaxation and sleepiness.
\par

\section{Discussion}
The analysis of multimodal interactions and parasocial attractions describes the common patterns used to induce ASMR experiences through audio-visual media. Our work depicts common ASMR interactions but does not contrast their effects with other online content. This section summarizes that the ASMRtists deliver ASMR effects through three experience patterns: multimodal social connection, relaxing physical intimacy, and sensory-rich activity observation. 

\subsection{ASMR as an Experience of Multimodal Social Connection} 
Prior research primarily examine ASMR triggers' characteristics, and their different physiological effects on the viewers \cite{PoerioMoreThanAFeeling, RichardBrainTingles, BarrattSensoryDeterminants}. Although ASMR videos are considered as a new pathway to connect creators and their viewers \cite{MaddoxASMRCommunity}, there is little knowledge regarding what specific interactions ASMRtists perform to best establish social connections. We find the social experience in YouTube ASMR videos is commonly offered and multi-modeled. Around 65\% of videos in our data contain the performer looking at the camera, which gives the illusion of a ``face-to-face'' interaction between the actor and the viewer, and 78\% of videos involve whispering. 70\% of ASMR videos have ASMRtists communicating verbally to the viewer, with 24\% of videos pretending that the performer can hear viewers' reply (talk-with videos). 59.07\% of videos used at least one type of touch interaction. The pervasive use of conversational content reveals that sound effects are not the only drivers of ASMR; ASMRtists engage viewers and induce ASMR through experiences of one-sided social connection. These results are consistent with the significance of face-to-face interactions noted in prior research \cite{StarrASMRSexuality}. In a multimodal conversation, the performer faces the viewer, communicates in whispers, touches the viewers through camera reaching, introduces and manipulates triggers, and emulates imagined scenarios. Since ASMR needs to be triggered with appropriate stimuli \cite{BarrattASMRFlowLike}, and because not all triggers ``work'' for all viewers, the diverse modalities allow viewers to try out and encounter triggers that can bring ASMR sensations. The social interactions could also foster the feeling of co-presence with the ASMR performer \cite{ZappavignaASMRRolePlay}. Interaction modalities such as whispering with/to the viewers and being spatially close up to the camera lead viewers to write more about the social processes in the comments. Even in audio-only videos without visual presentations, ASMRtists play fantasy and romantic roles and chat with the viewers in stories. The analysis of viewer comments suggests that viewers tend to leave more intimate comments to videos with those ASMR components. 
\par
These findings imply new pathways to design parasocial experiences with ASMR effects. ASMR interaction techniques can provide social exposure that increases closeness in asynchronous video communication. Video-based technologies incorporating ASMR effects and multimodal ASMR interactions may augment parasocial connection experiences. Since ASMR may offer positive affect, as well as intimate and relaxing experiences for viewers \cite{AndersenShiveriesASMR, PoerioInsomniaRelaxation}, face-to-face video communications can leverage ASMR interactions to transfer the process of speaking-listening to a richer experience with tingling sensations. The social experience pattern captured from ASMRtists' videos implies that technologies can incorporate ASMR interactions in multiple modalities such as whispering, camera-reaching, emulated back-and-forth conversation, and trigger manipulation in order to induce viewers' ASMR sensations. Users need both time and variety in order to see if tingles develop, and the multi-modalities allow for that temporal unfolding and variety. For example, applications such as video conferencing tools, podcasts, and social audio apps can potentially introduce multimodal ASMR to reduce the exhaustion and fatigue from long-time use \cite{NadlerZoomFatigue}. Voice-based virtual assistants \cite{ParvianenVoiceAssistants} may also include ASMR effects to reduce the robotic sound. 

\subsection{ASMR as an Experience of Relaxing Physical Intimacy}
Leveraging attraction and interaction techniques to demonstrate intimacy is also a typical pattern in ASMRtists' videos. Prior research has explored ASMR as an experience of digital intimacy \cite{AndersenShiveriesASMR, KlausenSafeandSound} as well as the ways ASMRtists create roleplay videos to foster intimate feelings with the viewers \cite{ZappavignaASMRRolePlay}. This paper overviews ASMRtists' techniques to design intimate experiences and how these techniques relax viewers and help with sleeping. We find that the most common camera shot type in ASMR videos is closeup -- framing the performer's face at a near distance while excluding most of their body. Around 30\% of videos have the ASMRtists pretending to touch the viewers through camera reaching. About 30\% of videos also make close-mic mouth sounds, and 12\% manipulate the microphone itself to simulate physical intimacy through sound interactions. These interactions are commonly performed in service-oriented videos such those involving massage, haircuts, makeup applications, etc. However, our comment analysis suggests that viewers do not express more intimacy to videos with intimate interactions than other videos. On the other hand, videos with close interactions have more comments regarding relaxation and sleepiness-related words. Our results imply that although ASMRtists virtually approach the viewers, viewers expressed relaxing and calming experiences more than intimacy to such videos. 
\par
Our findings suggest new opportunities to design ASMR-based applications to present intimacy and deliver soothing experiences. For example, ASMR interactions allow service providers such as masseurs and Reiki masters to offer virtual treatments through ASMR videos. This virtual therapy could provide a possible solution when in-person service is unavailable, or for users who cannot afford in-person treatment. People separated from loving relationships \cite{GanLoveAtADistance} or patients living in stressful hospital settings \cite{WallaceIntimacyHospitalSetting} need intimate interactions. ASMR effects with close-mic whispers and near-camera touching could potentially engender a feeling of intimacy to induce relaxing experiences. Virtual social encounters with ASMR performers could also provide alternatives for people with social difficulties (e.g., due to autism or social anxiety) to enjoy safe, calm, regularized social experiences on demand \cite{KampmannIntervensionSocialAnxiety}. To augment such experiences, designers can create new ASMR video interactions. For example, ASMRtists use the talk-with and camera-reaching techniques to mimic physical proximity. Novel interaction techniques such as VR, AR, and other telepresence technologies can be integrated to augment the social and virtual presence during ASMR videos. However, we want to remind the HCI community of the existance of sexual ASMR videos (S-ASMR) that are intentionally made for sexuality and sexual arousal \cite{StarrASMRSexuality}. The design for intimacy needs to differentiate ASMR videos from S-ASMR videos, especially to avoid young kids accessing an S-ASMR video without parental guidance.

\subsection{ASMR as an Experience of Sensory-rich Activity Observation}
Prior research have studied roleplay ASMRs as a primary type \cite{ManonASMR, ZappavignaASMRRolePlay, StarrASMRSexuality, LapinskaFixingYou, WeberFoodInfluenceSocialMedia}. However, our findings suggest that more than 70\% of videos in our data do not have roleplay scenarios. Also, in contrast to the wide use of social attraction and spatial proximity, most ASMRtists' videos do not use task attractions to elicit ASMR experiences. Only around 40\% of videos in our dataset have identifiable tasks and goals. These numbers indicate that intentional ASMR videos are not limited to roleplays; future work should include the diverse non-roleplays and taskless videos when examining ASMR performance and effects. Videos with tasks include the performance of various physical treatments, eating a large quantities of food, and displaying mundane activities such as playing cards or putting on makeup. Taskless videos can involve casual chatting or object or mic manipulation without showing the performer. The infrequent appearance in videos of purposeful tasks implies that ASMR effects do not require attention or real acts of care to take place. Therefore, many ASMRtists choose not to demonstrate abilities by completing tasks or making clear storylines, but instead remain focused exclusively on the production of triggering effects. The analysis of viewer comments further reveals that eating/drinking videos and videos without tasks or roleplays are associated with viewers' comments about the body and perceptual processes, indicating that these videos are prone to trigger bodily and perceptual experiences.
\par
Although personal attention, care, and intimacy are common elicitors of ASMR \cite{AndersenShiveriesASMR}, our findings suggest many ASMRtists also adopt the ``taskless'' activities in ASMR videos. Those videos don't particularly care about addressing viewers, adopting a stance of calculated indifference, and this disinvested attitude is designed to induce ASMR feelings. Therefore, videos that does not require close attention except for observing peaceful and repetitive activities -- videos such as crafting process demonstrations, instructions for applying makeups or skincare, and tutorials on organizing everyday objects -- may consider employing ASMR effects. Prior research suggested that ASMR is an ambient sensory effect in YouTube study-with-me videos \cite{LeeStudyWithMe}. Videos like these can potentially reduce the human presence and intentionally make tingling sounds in the background to trigger ASMR feelings. However, videos that include slow and dull tasks may evoke ASMR feelings unintentionally and could make viewers lose focus and feel sleepy. In those cases, ASMR may need to be avoided if the video is geared toward learning and requires attention. Designers may also consider conveying sensory-rich experiences through Mukbang ASMR or sound-focused ASMR. Watching food-eating videos has shown benefits to mitigate homesickness \cite{KellyHomesickness}. People watch Mukbang videos to gain multi-sensory immersion and ``commensality.'' \cite{TangMukbang} ASMR can be a sensory experience incorporated in human-food interaction \cite{DengHumanFoodInteraction, WeberFoodInfluenceSocialMedia}. Interaction designers can induce ASMR experiences by mouth and mic sounds to augment sensory pleasure. Technologies for sensory reality and relaxation (e.g., virtual reality for anxiety-related problems \cite{OingVRAnxiety}) can incorporate ASMR techniques such as eating/drinking sounds or sound-focused scenes to induce sense of presence and relaxing experiences.

\section{Conclusion and Future Work}
This work analyzed the multimodal interactions and parasocial experiences in 2663 YouTube ASMR videos. We annotated how ASMRtists use visual, sound, touch, taste, and roleplay triggers to deliver social, physical, and task attractiveness. We obtained the distribution of ASMR interaction modalities and parasocial attractions. The associations between interaction modalities and parasocial attractions reveal patterns of ASMR experiences. Feeling-oriented words were recognized from viewer comments in order to probe whether different ASMR interactions lead to different viewer feelings. Face-to-face orientation, whispering sounds, and touching objects are the most interaction modalities. Social interactions are common and multi-modeled. ASMRtists implement social attractions and virtually proximate the viewers, but most ASMR videos do not involve roleplays or contain purposeful tasks. Our results summarize that YouTube ASMR videos provide three experiences: multimodal social connection, relaxing physical intimacy, and sensory-rich activity observation. These experiential descriptions seek to foster future media productions on a wide array of platforms that include ASMR interactions and effects. 
\par
Moving forward, we hope this work serves as a seminal study to inspire more ASMR-augmented designs. There are also many open-ended questions to be addressed by HCI researchers and practitioners. First, one limitation of this work is that we only consider intentional ASMR created and shared by YouTube ASMRtists to induce ASMR experiences specifically. Prior studies noticed viewers also experience ASMR with videos such as Bob Ross' \textit{The Joy of Painting} and a recording of \textit{Lectures on Quantum Field Theory} \cite{ManonASMR, GallagherAetheticsOfASMR}, which are not made for ASMR but contains ASMR properties. We did not include unintentional videos without ``ASMR'' labels due to difficulties recognizing and collecting them from YouTube. We also consider intentional ASMR interactions to be purposefully designed and performed; therefore, easier to be adopted in design. Future research may compare and contrast the effects of the two ASMR video types. Second, this work does not interview actual viewers to obtain their in-situ feelings of ASMR interactions; viewers' reactions to different ASMR interactions were obtained from video comments. It is possible that viewers do not externalize all of their feelings of intimacy or relaxation in comments. However, we believe this work provides an overview of ASMR interaction techniques that can guide future studies to examine ASMR-based intimacy and well-being in various use cases \cite{LomanowskaOnlineIntimacy}. Future research needs to assess the actual effects of ASMR interactions of different people and in different contexts, especially when ASMR interactions are designed for people with social anxiety or disabilities. Third, YouTube creators contribute vernacular creativity \cite{Burgess2012YouTubeMedia} to build parasocial relationships. HCI researchers should consider interviewing ASMRtists or involving them in participatory design to understand their preferences and difficulties in managing parasocial interactions. Last, the growing ASMR communities across different cultures \cite{TangMukbang, LeeStudyWithMe} encourages HCI studies to examine how ASMR videos affect the creator-viewer communications and relationships. It is valuable to expand ASMR research to non-English videos to have a cross-cultural understanding of ASMR.
\bibliographystyle{ACM-Reference-Format}
\bibliography{references.bib}
\appendix

\section{Data Protection Impact Assessment (DPIA)}\label{dpia}
 All videos are publicly shared when gathered, and the researchers did not directly interact with any ASRMtists to collect any information about the video creators. We acknowledge this work uses automated data processing tools (LIWC and PMI) to recognize behavioral information from YouTube comments. To protect YouTube users' privacy, we de-linked the YouTube account information from the collected comments after the data was collected. To our best knowledge, YouTube does not support searching comments. LIWC and PMI are lexicon-based techniques that count the appearance of words related to the measured dimensions. Only words related to common feelings such as social process, intimacy, perception, and relaxation are quantified. We did not intentionally recognize or process any identity information.
 \par
 The data annotation on Amazon Mechanical Turk only involves identifying objective information from the video (e.g., how the ASMRtist interacts with trigger objects or the camera). All questions are multi-choice questions with pre-defined categories. We do not ask participants to provide personal information or subjectively describe any video content. To protect the participants who are uncomfortable with ASMR videos, all participants must pass a qualification test before the study. The participant must indicate that they had watched ASMR videos before and did not feel them disturbing. At the beginning of the annotation, we also informed that if the video content makes the participant uncomfortable, they should close the survey immediately. With the careful research steps, the IRB office at the authors' institute granted an exemption to the protocol of this study.
\end{document}